\documentclass[a4paper,11pt]{article}
\usepackage[utf8]{inputenc}
\usepackage{caption}
\usepackage{subcaption}
\usepackage{jcappub}
\usepackage{aas_macros}
\usepackage{rotating}
\usepackage{lineno}
\title{The cosmic-ray positron excess and its imprint in the Galactic gamma-ray sky}

\author[a]{M. Rocamora}
\author[b,c]{, Y. Ascasibar}
\author[b,d]{, M. A. S\'{a}nchez-Conde}
\author[e,f]{, M. Wechakama}
\author[b, d, g]{, P. de la Torre Luque}

\affiliation[a]{Universität Innsbruck, Institut für Astro- und Teilchenphysik, Technikerstr. 25/8, 6020 Innsbruck, Austria}
\affiliation[b]{Departamento de F\'{i}sica Te\'{o}rica, M-15, Universidad Aut\'{o}noma de Madrid, E-28049 Madrid, Spain}
\affiliation[c]{Centro de Investigación Avanzada en Física Fundamental (CIAFF-UAM), Madrid, Spain}
\affiliation[d]{Instituto de F\'{i}sica Te\'{o}rica UAM-CSIC, Universidad Aut\'{o}noma de Madrid, C/ Nicol\'{a}s Cabrera, 13-15, 28049 Madrid, Spain}
\affiliation[e]{Kasetsart University, 50 Ngamwongwan Rd., Lat Yao, Chatuchak, Bangkok 10900, Thailand}
\affiliation[f]{National Astronomical Research Institute of Thailand, Don Kaeo, MaeRim, Chiang Mai 50180, Thailand}
\affiliation[g]{Stockholm University and The Oskar Klein Center for Cosmoparticle Physics, Alba Nova 10691 Stockholm, Sweden}

\emailAdd{manuel.rocamora@uibk.ac.at}
\abstract{We study the origin of the positron excess observed in the local cosmic-ray spectrum at high energies, and relate it to the cosmic rays and gamma-ray emission across the entire Galaxy.
In particular, we explore the hypothesis of a single, dominant source accountable for primary electron-positron pairs. 
Since we are agnostic about the physical nature of the underlying source population, we consider four simple models that are representative of young pulsars, old stars (as a tracer of millisecond pulsars), and annihilating dark matter particles. In the dark matter hypothesis, we consider both a cored and a cuspy model for the halo in the Milky Way.
Then, we compare the associated gamma-ray sky maps with \textit{Fermi}-LAT data.
The aim of this work is not to derive constraints or upper limits for the different models considered, but rather to explore the possibility, as a proof of concept, of building a self-consistent model able to explain simultaneously the origin of all cosmic-ray species, including positrons, as well as the Galactic center GeV gamma-ray emission.
We find that the emission arising from pulsar wind nebulae is fairly concentrated near the mid plane, and therefore additional cosmic-ray sources must be invoked to explain the emission at the center of the Galaxy.
If the local positron excess were mainly due to millisecond pulsars, inverse Compton scattering by the particles injected in the Milky Way bulge would naturally account for a non-negligible fraction of the central gamma-ray emission.
The case of annihilating dark matter is very sensitive to the precise shape of the dark matter profile. The results obtained for a standard NFW cuspy profile are above the gamma-ray measurements by as much as a factor of 2 in some regions of the Galaxy, while the results for an isothermal, cored profile are still compatible with the data. However, the cross-sections exceed the current constraints.
}
\begin{document}
\maketitle
\flushbottom

\section{Introduction}

%% -- cosmic rays --
 
After more than a century of intense research, the origin of cosmic-ray (CR) positrons is still not fully understood.
It is now widely accepted that most Galactic CRs are accelerated in supernova remnants \cite{Blasi_2013}.
Secondary positrons, produced by the collision of the relativistic protons with the surrounding interstellar matter, are able to account for the observed spectrum of Galactic positrons up to $\sim 10$~GeV \cite{Moskalenko_1998}.
However, observational evidence strongly suggests the presence of an additional Galactic source yet to be identified. Positrons lose their energy through a variety of processes, until they eventually annihilate with electrons in the interstellar medium. The morphology of the 511 keV line (see, e.g.,~\cite{Prantzos_2011}, and references therein) exhibits an unexpected excess near the Galactic centre. This excess is more consistent with the morphology of the Milky Way bulge than with the distribution of the supernova remnants, which traces more closely the Galactic disk (see, e.g., Refs.~\cite{Vincent_2012, luque2023multimessenger}). Moreover, the PAMELA experiment measured in 2008 the energy spectrum of different CR species between 1 and 100 GeV \citep{PAMELA}.
While the results obtained for the positron abundance at energies below 10 GeV were fairly compatible with the predictions of the theoretical models based on a secondary origin (i.e. from primary proton spallation), it became clear that extra sources of primary positrons were necessary to explain the observed positron fraction above that energy.
These results were later confirmed by other experiments, such as AMS-02 \cite{AMS_positrons} and the \textit{Fermi} Large Area Telescope (\textit{Fermi} LAT) \cite{FERMI_positrons}, and a secondary origin of the high-energy positrons has been confidently ruled out by comparison with the antiproton spectrum \cite{Diesing_2020}.
The nature of the required source of primary positrons is nonetheless heavily debated, and several theories have been proposed.
Currently, the most favoured candidates are Pulsar Wind Nebulae (PWN) \cite{Hooper_2009, Hooper_2017, Cholis_2018, DiMauro_2019, Manconi_2020, Fornieri_2020, Evoli_2021} and the annihilation of Dark Matter (DM) particles within the Milky Way halo \cite{Boudaud_2015, Jin_2015, Jin_2020, Zhan_2023}. These two sources have been extensively studied, focusing on reproducing the local positron spectrum without discussing the implications beyond the Solar neighbourhood in detail.

%% -- gamma rays --

A complementary view of the CR population arises from the observation of their gamma-ray radiation.
The measurements conducted by \textit{Fermi} LAT have significantly enhanced our comprehension of the high-energy sky in the past 15 years. However, there are some features of the diffuse emission whose origin is still an open question.
The GeV excess around the centre of the Milky Way, discovered in 2009 \cite{Goodenough_2009,Hooper_2011a,Hooper_2011b,Boyarsky_2011,Abazajian_2012,Gordon_2013,Abazajian_2014,Calore_2015a,Calore_2015b,Daylan_2016,Ajello_2016,Ackermann_2017,DiMauro_2019_gev}, was soon reported to be consistent with DM annihilation on the basis of both its spectrum and its spherically-symmetric spatial morphology.
Subsequent studies proposed a population of re-accelerated MilliSecond Pulsars (MSPs) as an alternative explanation \cite{Bartels_2015, Bartels_2018}.
These sources would also inject positrons throughout the whole Galaxy, and therefore they have also been considered as possible contributors to the local positron excess \cite{Venter_2015, Linares_2021}.

Many previous studies of the gamma-ray sky describe it in terms of spatial and spectral templates that trace different components \citep[e.g.][]{Acero_2016, Storm_2017, Pohl_2022}, while more physically-motivated models of the injection and propagation of CR do not specifically address the positron excess problem \cite{Ackermann_2012,Orlando_2018,Cholis_2022, Thaler_2023, DeLaTorre_2023}.
%% -- this work --
In this work, we test whether the contribution of DM, PWN or MSP could account for both the high-energy positrons observed in the Solar neighbourhood, as well as the gamma-ray emission from the whole Galaxy.
The aim of this work is not to derive constraints or upper limits to the different possible models here considered, but to explore the possibility, as a proof of concept, of building a self-consistent model that is able to explain simultaneously the origin of all CR species, including positrons, as well as 100\% of the gamma-ray emission.
Our CR propagation model is fully described in Section~\ref{sec:model}.
We follow a model-agnostic approach, where the positron energy spectrum is set from a fit of the local measurements \cite{Trotta_2011, Korsmeier_2016, Evoli_2019, Evoli_2020, Jin_2020, Luque_2021}, and the spatial distribution depends on the nature of the sources tested in this work.
The results of these fits are presented in Section~\ref{sec:local_CR}.
Section~\ref{sec:gamma} focuses on the comparison between the predicted gamma-ray emission and the all-sky maps from the \textit{Fermi}-LAT Collaboration\footnote{\textit{Fermi}-LAT data repository: \url{https://fermi.gsfc.nasa.gov/ssc/data/access/}}.
The implications for the DM, PWN, and MSP scenarios are discussed in Section~\ref{sec:discussion}, and our main conclusions are briefly summarised in Section~\ref{sec:conclusions}.

%% -- MODEL --
\section{Cosmic-ray propagation model}
\label{sec:model}

We assume that the injection and propagation of cosmic rays in the Milky Way is governed by the stationary advection-diffusion equation \cite{Ginzburg_64,Berezinskii_90}:
\begin{align}
   \nonumber -\overrightarrow{\nabla}\cdot\left(D\overrightarrow{\nabla}\Psi_i + \overrightarrow{v}_w\Psi_i\right) + \frac{\partial}{\partial p}\left[p^2 D_{pp}\frac{\partial}{\partial p} \left(\frac{\Psi_i}{p^2}\right)\right] - \frac{\partial}{\partial p} \left[\dot{p}\Psi_i - \frac{p}{3}\left(\overrightarrow{\nabla}\cdot \overrightarrow{v}_w\right)\Psi_i\right] =\\ Q_i + \sum_{i<j}\left(c~\beta~n_{gas}~\sigma_{j\rightarrow i} + \frac{1}{\gamma \tau_{j\rightarrow i}}\right)\Psi_j - \left(c~\beta~n_{gas}~\sigma_i + \frac{1}{\gamma\tau_i}\right)\Psi_i
   \label{eq:prop}
\end{align}
where $\Psi_i$ denotes the spectrum (number of particles per unit volume per unit momentum) of CR species $i$, $D$ is the spatial diffusion coefficient, and $v_w$ is the effective wind velocity in the Galaxy;
$p$ denotes the momentum of the particles, and $D_{pp}$ refers to diffusion in momentum space; the $\dot{p}$ and the following term represent the momentum and the adiabatic losses, respectively; $Q_i$ is the source term (injection spectrum per unit time);
$c$ represents the speed of light, $\beta$ and $\gamma$ are the velocity and Lorentz factor, respectively, $n_{gas}$ is the number density of the interstellar gas particles, $\sigma_{j\rightarrow i}$ is the spallation cross-section of species $j$ into $i$ by collision with gas particles, $\tau_{j\rightarrow i}$ represents the lifetime for spontaneous decay, $\sigma_i \equiv \sum_{j<i} \sigma_{i\rightarrow j}$, and $1/\tau_i \equiv \sum_{j<i} 1/\tau_{i\rightarrow j}$.

This equation is solved numerically using the DRAGON CR propagation code\footnote{\url{https://github.com/cosmicrays/DRAGON2-Beta_version}} \citep{DRAGON}
on a uniform spatial and energy grid.
We use a simulation box given by a 2-dimensional cylindrical spatial grid with $R \in [0, 12]$~kpc and 121 grid points, and $z \in [-18, 18]$~kpc with 361 grid points, i.e. a uniform spatial resolution of 0.1 kpc in both dimensions, and study energies in the range $E_{k} \in\rm [10~MeV, 100~TeV]$, using 90 logarithmically spaced energy bins.
Several tests have been carried out with different configurations to ensure that the simulation grid has sufficient resolution near the Galactic centre and that it extends sufficiently far away from the mid-plane. We have verified that our results do not depend on these numerical parameters if we increase the number of grid points and/or the maximum values of $R$ and $z$ beyond the adopted values.

\subsection{Source terms}
\label{sec:source}

Following previous works \citep[e.g][]{Fornieri_2020, fornieri_2021, sync_spec}, we use a broken power-law
\begin{equation}
    Q_{\rm SNR}(R, z, E) = Q_0\ S(R, z)
    \begin{cases}
    \left(\frac{E}{E_1}\right)^{\alpha_0} \iff E<E_1\\
    \left(\frac{E}{E_1}\right)^{\alpha_1} \iff E_1<E<E_2\\
    \left(\frac{E_2}{E_1}\right)^{\alpha_1}\left(\frac{E}{E_2}\right)^{\alpha_2} \iff E>E_2\\
    \end{cases}
    \label{eq:injection}
\end{equation}
to model the injection of protons and heavier nuclei by SuperNova Remnant (SNR) shocks.
For electrons, an additional exponential cut-off $e^{-E/E_{cut}}$ is included at high energies.
The overall normalisation constant $Q_0 \equiv Q(R_\odot, 0, E_0)$, i.e. the injection rate at the Solar location ($R=R_\odot=8.3$~kpc, $z=0$) at an arbitrary reference energy $E_{0} \equiv 1$~GeV, as well as the logarithmic slopes $\alpha_0$, $\alpha_1$, $\alpha_2$, and the break energies $E_1$ and $E_2$, are free parameters to be fitted independently for each particle species, whose values are set to reproduce the local CR data.

In order to explain the observed positron spectrum, it is necessary to postulate a new source of primary electron-positron pairs, which we refer to as the `extra component'.
To model its injection spectrum as a function of energy, we use a single power-law with an exponential cut-off \citep{AMS_comp}:
\begin{equation}
    Q_{\rm extra}(R, z, E) = Q_0\ S(R, z) \left(\frac{E}{E_0}\right)^{\alpha} e^{-E/E_{\rm cut}},
    \label{eq:inj_extra}
\end{equation}
where the normalisation $Q_0$, the logarithmic slope $\alpha$ and the cut-off energy $E_{\rm cut}$ are varied as free parameters, while $E_0 \equiv 1$~GeV merely denotes, once again, the reference energy adopted to express $Q_0$.
Although every single pulsar (PWN or MSP) may feature a unique injection spectrum, this simple analytical form provides a fair approximation to describe the sum over the whole population of individual sources.
In the context of DM annihilation, instead of assuming a specific particle physics model (e.g. mass and branching ratios for the different annihilation channels), we here remain completely agnostic about the precise nature of the DM particles, and simply use this phenomenological injection spectrum under the assumption that their annihilation is the main physical mechanism responsible for the local positron excess. The spectrum of particles produced by DM annihilation could feature sharp cut-offs. However, we will limit ourselves to study simple exponential ones.

\begin{figure}
    \centering
    \begin{subfigure}{\textwidth}
        \includegraphics[width = \textwidth]{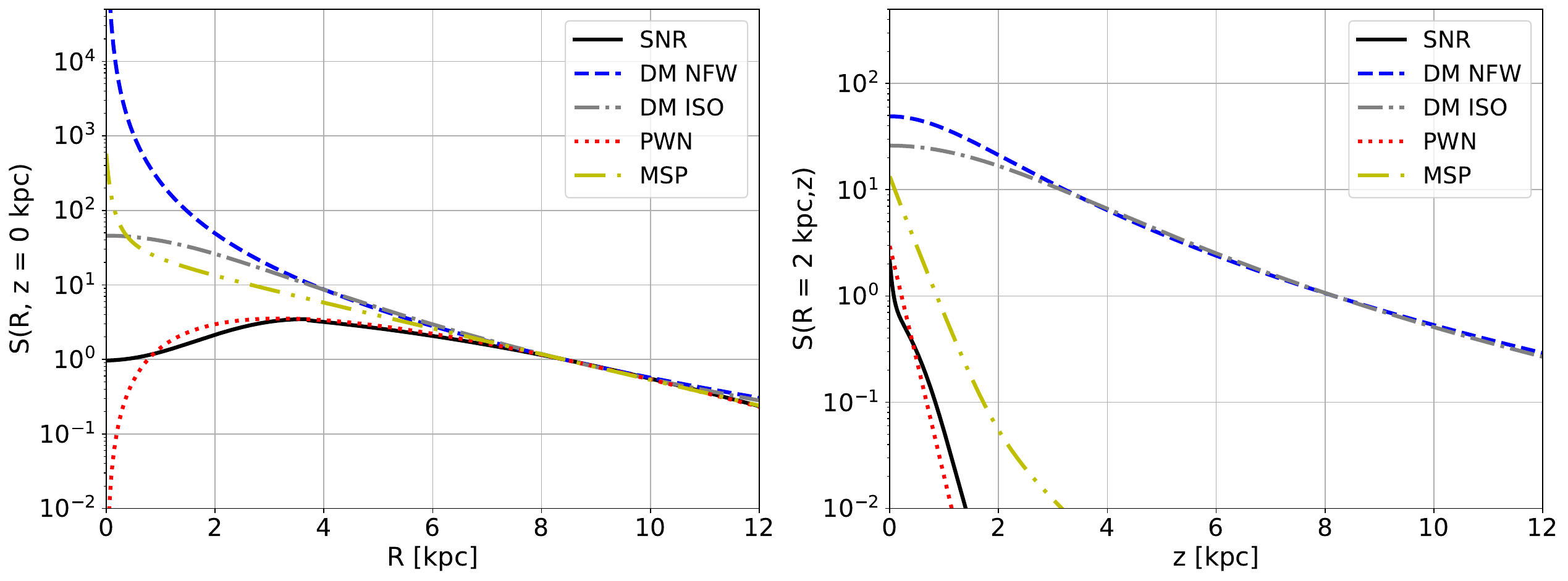}
    \end{subfigure}
    \caption{Particle injection profiles as a function of Galactocentric radius at z = 0 (left) and in terms of height from the Galactic plane at R = 2 kpc (right).}
    \label{fig:profiles}
\end{figure}

Regarding the spatial distribution $S(R, z)$,
%normalised to the Solar neighbourhood, i.e. $S(R_\odot, 0) \equiv 1$,
we adopt a Ferriere \citep{Ferriere} profile:
\begin{align}
    \nonumber
    S_{\rm SNR}(R, z) = 7.3 \exp\left(-\frac{R-R_\odot}{4.5} - \frac{|z|}{0.325}\right)\ + & \\
    + \left[0.79 \exp\left(-\left(\frac{z}{0.212}\right)^2\right) + 0.21\exp\left(-\left(\frac{z}{0.636}\right)^2\right) \right] \times
    &
    \begin{cases}
        177.5 \exp \left(-\frac{(R-3.7)^2}{2.1^2}\right) \iff R < 3.7 \\
        50.0 \exp \left(-\frac{R^2-R_\odot^2}{6.8^2}\right) \iff R > 3.7
    \end{cases}
\end{align}
to describe the distribution of Galactic SNR.

For PWN, we adopt a Lorimer profile \cite{Lorimer}:
\begin{equation}
    S_{\rm PWN}(R, z) = \left(\frac{R}{R_\odot}\right)^{1.9} \exp\left(-5.0\, \frac{R-R_\odot}{R_\odot} - \frac{|z|}{0.2}\right)
\end{equation}
to describe the distribution of pulsars in the Milky Way.

The only MSPs capable of accelerating particles above 10 GeV are those inside compact binary systems \cite{Venter_2015}.
A limited number of these systems have been discovered in the Galaxy so far, and so their distribution is poorly known. Therefore, we decide to use a McMillan profile in this case: \cite{McMillan}
\begin{align}
    \nonumber
    S_{\rm MSP}(R, z) = \frac{98.4}{\left(1+R'/0.075\right)^{1.8}}\exp\left(- \frac{R^2 + (z/0.5)^2}{2.1^2} \right) + \frac{896}{2\cdot300} \exp\left(- \frac{|z|}{0.3} - \frac{R}{2.5}\right) + \\ + \frac{183}{2\cdot900} \exp\left(- \frac{|z|}{0.9} - \frac{R}{3.02}\right),
\end{align}
where $R' = \sqrt{R^2 + (z/0.5)^2}$.
This function describes the overall distribution of stars within the Galaxy, as a first order approximation of the location of MSPs.

Finally, the annihilation rate of DM particles in the Milky Way halo is proportional to the square of the DM density, whose behaviour in the innermost regions is still debated (the so-called ``cusp-core'' problem \cite{FloresPrimack94,Moore94,DiCintio_2014,Benito_2019}).
In this work, we will study two distributions: the one proposed in Ref.~\cite{NFW} following the results found in cosmological simulations (the so-called 'NFW' profile), and an isothermal (ISO) profile \cite{Begeman_1991}. These are representative of both cuspy and cored DM distributions, respectively. The difference between cuspy and cored profiles lie at the center of the DM halo, while the former has a pronounced peak, the latter is flat. The NFW and ISO profiles squared read as follows:
\begin{equation}
    S_{\rm NFW}(r) = \frac{1}{(r/r_s)^2(1+r/r_s)^4}
\end{equation}
\begin{equation}
    S_{\rm ISO}(r) = \frac{1}{(1+(r/r_{iso})^2)^2}
\end{equation}
where $r \equiv \sqrt{R^2+z^2}$ is the spherical radial coordinate, whereas $r_s = 20$~kpc \cite{Pato_2015} and $r_{iso} = 3.5$~kpc \cite{Bertone_2005} denote the scale radius of the DM halo in each profile.

The spatial distributions of the different CR sources are shown in Figure~\ref{fig:profiles}.
The density of SNR and PWN is quite uniform across the inner Milky Way disk, peaking a few kpc away from the Galactic centre, whereas MSP and DM are much more concentrated, tracing the morphology of the stellar bulge and of the DM halo, respectively.
Therefore, they are expected to yield very different predictions about the CR and gamma-ray spectra in the innermost regions of the Galaxy, once they are fit to the Solar neighbourhood measurements.
As one can see on the right panel, the DM profiles (both NFW and ISO) extend up to much higher latitudes that the exponentially decaying injection profiles of the pulsar scenario (PWN and MSP).

\subsection{Propagation parameters}

For the diffusion coefficient, we assume that it follows a single power-law in particle rigidity $\rho$, and it increases exponentially with the height $|z|$ over the Galactic mid-plane:
\begin{equation}
    D(\rho, z) = D_0 \left(\frac{\rho}{\rho_0}\right)^\delta \exp(|z|/z_t),
    \label{eq:diff}
\end{equation}
with reference rigidity $\rho_0$ = 1 GV.
The normalisation $D_0$, the logarithmic slope $\delta$, and the characteristic scale height $z_t$ are to be fitted.
This leads to a diffusion in momentum space, also known as re-acceleration, whose coefficient is given by \cite{Berezinskii_90}:
\begin{equation}
    D_{pp} = \frac{4}{3\delta (4-\delta^2)(4-\delta)}\frac{p^2v_A^2}{\langle D\rangle},
\end{equation}
where $v_A$ is the Alfvén velocity and $\langle D\rangle$ is the spatial diffusion coefficient from Eq.~\ref{eq:diff} averaged over all the directions.

The energy losses considered in this work are pion production, Coulomb scattering, ionization, bremsstrahlung, inverse Compton scattering and synchrotron radiation, all of them included by default in DRAGON.
For the magnetic field, we use the model of \cite{Pshirkov_2011}, adding a turbulent component to match the observed local value of $\sim 6$ $\mu G$ \cite{Ferriere}. We assume the shape of the turbulent component as
\begin{equation}
    B_{turb} = B_{0,turb} \; \exp\left(-\frac{R-r_\odot}{r_0}\right) \exp\left(-|z|/z_t\right),
\end{equation}%poner bien la ecuacion
with $B_{0,turb}$ = 5.5 $\mu$G, $r_0$ = 8.5 kpc and same $z_t$ as for the diffusion coefficient.

In order to ensure self-consistency, we adopt the same gas density model as the HERMES\footnote{\url{https://github.com/cosmicrays/hermes}} code \cite{HERMES}, that we will later use to compute the gamma-ray emission.
This model is composed of 11 Galactocentric rings of gas column density derived from the HI4PI survey \citep{HI_survey} and the CfA survey \citep{H2_survey} for atomic and molecular gas, respectively.
For the molecular gas, we use an $X_{CO}$ factor of $1.9 \times 10^{20}$~cm$^{-2}$~K$^{-1}$/(km~s$^{-1}$) \cite{Strong_1996}.

We make use of the DRAGON default cross sections, described in \cite{DRAGON_xsec}, to calculate the nuclear secondaries, and the Kamae cross section \cite{Kamae_2006} to calculate the secondary electrons and positrons.

%-- LOCAL CR --
\section{Cosmic-ray spectra}
\label{sec:local_CR}

In order to determine the free parameters in our model, we make use of $B/C$, $He$, $p$, $e^-$ and $e^+$ data from AMS \citep{AMS}, as well as H.E.S.S. measurements of the combined $e^- + e^+$ spectrum \cite{Aharonian_2008}.
We also consider the Voyager-1 mission data \citep{Voyager} to investigate the effect of the Sun at low energies, that we describe by means of the Force Field approximation \cite{Usoskin_2005, Usoskin_2011} for simplicity, where the modulation potential $\langle\phi_{mod}\rangle$ is an additional free variable.
This description has large uncertainties, especially at low energies, but a more detailed modelling of the Heliosphere is beyond the scope of this work.

Any physically motivated model of the Galactic CRs necessarily involves a large number of free parameters.
In our case, we have six (normalisation, logarithmic slopes, and breaks) for the primary injection of each particle species (electrons, protons, Helium, Carbon, and Oxygen) by SNR, another three (normalisation, slope, and cut-off) for the extra component (be it PWN, MSP, or DM), and five more associated to particle propagation.
Our prior knowledge of their true values is, at best, uncertain, and they are strongly degenerate.
Exploring in a proper way such a vast parameter space with $(6 \times 5 + 3 \times 3 + 5) = 44$ dimensions and evaluating the uncertainties associated to our fits (with, e.g. a Markov-Chain Monte Carlo approach) is extremely challenging from a computational point of view.
Moreover, it would only provide a rigorous quantification of the statistical uncertainties, yet neglecting the significant systematic biases incurred by our choice of astrophysical parameters associated to the Milky Way, such as the gas density, magnetic field, radiation field, and spatial distribution $S_i(R, z)$ of the different sources. 

For these reasons, we do not attempt to find the model that best fits the observational data.
Instead, this work is intended as a proof of concept, aiming to investigate whether it is possible to explain all the available CR and gamma-ray measurements in terms of two source populations, namely SNR and the `extra component', and whether PWN, MSP, or DM would be viable candidates for the latter.
To that end, we adjust manually the parameter values until we find a $\sim1\sigma$ agreement with the data. 

\subsection{Fitting procedure and baseline model}
We first set the CR propagation parameters and the primary injection of nuclei in the baseline SNR model.
Then, we fit the injection spectrum of primary electrons by SNR and the cross-section for secondary positron production.
Finally, we investigate and discuss the injection of additional primary electrons and positrons in our four `extra component' scenarios.
During the fitting procedure, we could not explain the low energy tail of the positron spectrum by only varying the propagation parameters of the model, with a flux prediction always around a factor of 2 below the data. Because of that, and given the uncertainties in the modelling of the local flux of secondary positrons \cite{DelaTorreLuque:2023zyd, Koldobskiy_2021,Orusa_2022}, we apply a fudge multiplicative factor, $k_{\sigma}$ to the cross section. 
This parameter, constant with the energy, simply scales the cross section. It is included in our fitting procedure and allows us to reproduce the low energy positrons.

\begin{figure}
    \includegraphics[width=\textwidth]{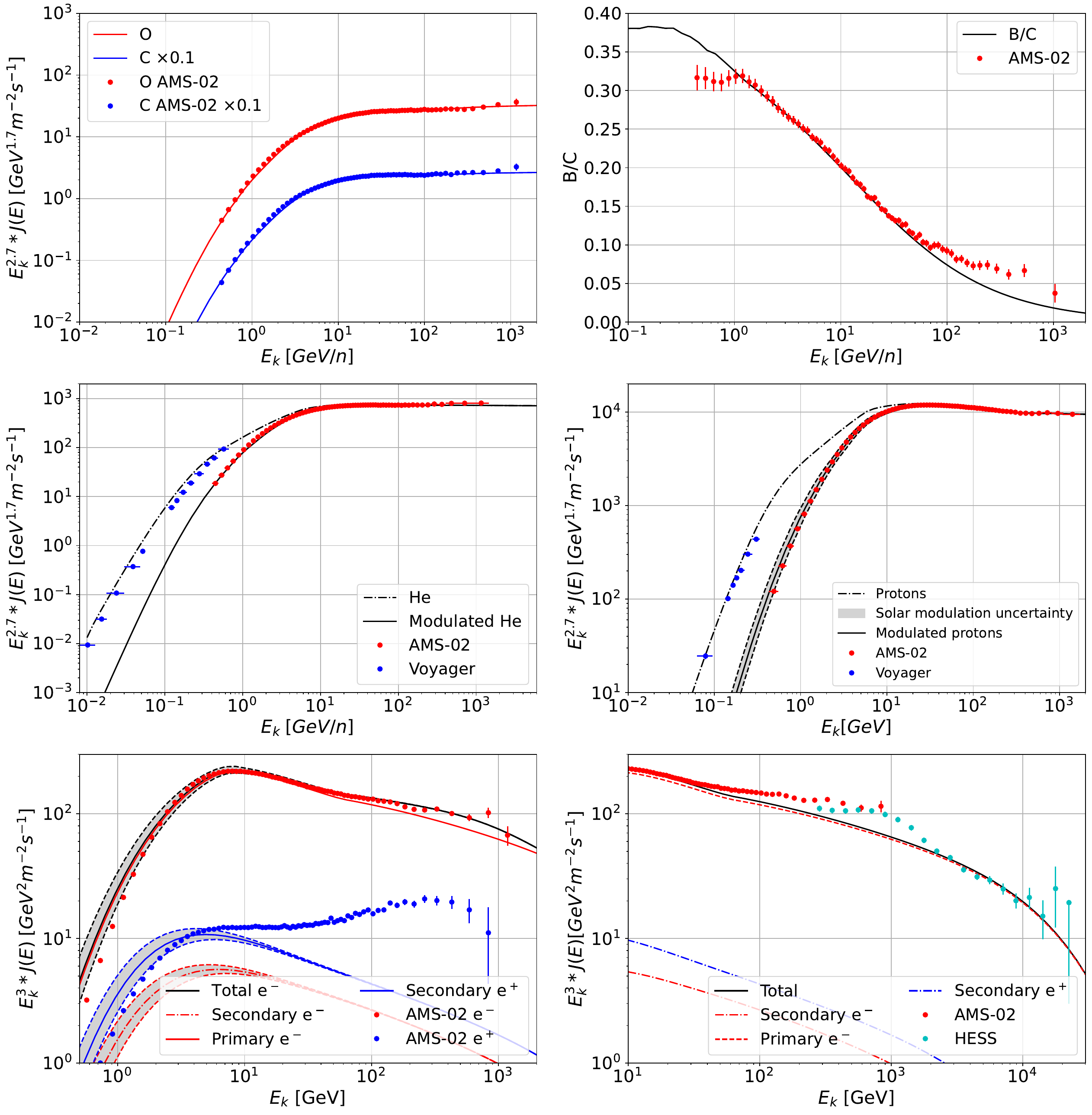}
    \caption{CR spectrum of our best-fit model at Earth for Oxygen and Carbon, the latter is multiplied by 0.1 for a better visibility (top left); Boron-to-Carbon ratio (top right); Helium (middle left); protons (middle right); electrons and positrons at low energies (bottom left); and all-leptons at high energies (bottom right), compared to data from either AMS-02 \cite{AMS}, Voyager \cite{Voyager}, or HESS \cite{Aharonian_2008}. The nuclei fluxes are multiplied by E$^{2.7}$, while the CR leptons by E$^3$. For Helium and protons, we show the effect of the solar modulation by showing AMS data (red) and Voyager data (blue). For the low-energy leptons, the red and blue colors stand for e$^-$ and e$^+$, respectively. The dashed lines and grey bands show the solar modulation uncertainty.}
    \label{fig:crnoextra}
\end{figure}

\begin{table}
    \centering
    \begin{tabular}{|c|c|c|c|c|c|}
        \hline
        $D_0$ [cm$^2$/s] & $\delta$ & $v_{\rm A}$ [km/s] & $z_{\rm t}$ [kpc] & $\langle\phi_{\rm mod}\rangle$ & $k_{\sigma}$ \\ \hline
        $1.1 \times 10^{28}$ & $0.68$ & $13.0$ & $4$ & $0.58$ & 1.5 \\ \hline
    \end{tabular}
    \caption{Propagation parameters of our model, manually set by comparison with local measurements of the CR spectra.}
    \label{tab:prop}
\end{table}

\begin{table}
    \centering
    \begin{tabular}{|c|c|c|c|c|c|c|c|c|}
        \hline
        Species & $Q_0$ & $-\alpha_0$ & $E_1$ & $-\alpha_1$ & $E_2$ & $-\alpha_2$ & $E_{\rm cut}$ \\
         & [GeV$^{-1}$ m$^{-3}$ Myr$^{-1}$] & & [GeV] & & [GeV] & & [GeV] \\ \hline
        O & $1.72 \times 10^{-7}$ & 2.00 & 7.0 & 2.25 & 100 & 2.03 & - \\ \hline
        C & $1.33 \times 10^{-7}$ & 2.00 & 7.0 & 2.23 & 100 & 2.03 & - \\ \hline
        He & $3.62 \times 10^{-6}$ & 2.00 & 7.0 & 2.21 & 100 & 2.05 & - \\ \hline
        H & $2.17 \times 10^{-5}$ & 1.80 & 7.0 & 2.21 & 335 & 2.05 & - \\ \hline
        $e^-$ & $7.91 \times10^{-7}$ & 2.00 & 8.0 & 2.70 & 60 & 2.46 & $2 \times 10^4$ \\ \hline
    \end{tabular}
    \caption{Injection parameters of the baseline SNR model. $Q_0$ is given at $E_0$ = 1 GeV.} 
    \label{tab:inj}
\end{table}

In the baseline model, all CRs are injected by sources following the distribution of supernova remnants.
The proposed values of the propagation parameters are quoted in Table~\ref{tab:prop}, whereas the injection of each particle species is specified in Table~\ref{tab:inj}.
The predictions for the CR flux, J(E), of this model are compared in Figure~\ref{fig:crnoextra} with the available observational measurements.
Although some discrepancies are still present, one may conclude that, in general, all the hadronic spectra (H, He, B/C, and O) are fairly well reproduced, taking into account current systematic uncertainties.

The most notable deviation corresponds to the Boron-to-Carbon ratio, both at low and high energies.
We find that above $\sim100$~GeV this ratio starts to deviate more significantly from the data. This has been explained in the literature as a change in the diffusion regime~\cite{Aloisio_2013,Evoli_2019}. However, we do not model such a feature because it introduces more degenerate free parameters in our procedure. In our case, where such a break is not modelled, we note that it is necessary to introduce a similar feature in the injection spectra \cite{Boschini_2020} at the energy scale we label as $E_2$.
For Oxygen, Carbon and Helium, we adopt $E_2 = 100$~GeV, and a somewhat larger value is required in order to accurately reproduce the observed proton spectrum, that displays a significant feature at $E_2 \sim 300$~GeV.
In the intermediate energy range $E_1 < E < E_2$, our logarithmic slope $\alpha_1$ is slightly steeper than the canonical value $\alpha \sim -2$.

In what concerns the propagation parameters, that are mainly determined by the secondary species B and e$^+$, we observe that the resulting diffusion index $\delta$ = 0.68 is a bit above the values typically obtained from the quasi-linear theory of diffusion, which predicts an index ranging between 1/3 and 1/2, depending on the turbulence model \cite{Blasi_2012}. Nevertheless, recent studies have also given results consistent with ours \citep{Evoli_2019, Evoli_2021, Reichherzer_2022, DiMauro_2023}. 
The Alfvén velocity varies a lot among the different studies, but it is typically lower than 40~km/s \cite{Trotta_2011, Evoli_2020, Fornieri_2020, Luque_2021, Thaler_2023}, which is consistent with our result of 13~km/s.
A height $z_t = 4$~kpc has been typically used and is in agreement with current observations of the $^{10}$Be isotope \cite{Evoli_2019, Evoli_2020}.
Note that in our model this is not a sharp boundary, but a representative scale height of the Galactic CR halo, and we stress the importance of simulating a much larger box in the vertical direction in order to achieve numerical convergence in the particle spectra, as well as to properly compute the gamma-ray emission at high Galactic latitudes.
Finally, the modulation potential is also in good agreement with the results of ground-based measurements \cite{Usoskin_2005, Usoskin_2011}, and we consider that our fudge multiplicative factor for the positron production cross-section, $k_{\sigma}$, is compatible with the present uncertainties~\cite{DelaTorreLuque:2023zyd}.

\subsection{Extra component}

\begin{table}
    \centering
    \begin{tabular}{|c|c|c|c|c|}
        \hline
        %Model & $E_0$ [GeV] & $Q_0$ [GeV$^{-1}$ m$^{-2}$ s$^{-1}$ sr$^{-1}$] & $-\alpha$ & $E_{\rm cut}$ [GeV] \\ \hline
        Model & $Q_0$ & $-\alpha$ & $E_{\rm cut}$ \\
         & [GeV$^{-1}$ m$^{-3}$ Myr$^{-1}$] & & [GeV] \\ \hline
        DM NFW & $1.22 \times 10^{-10}$ & 1.30 & 900 \\ \hline
        DM ISO & $1.22\times10^{-10}$ & 1.30 & 900 \\ \hline
        PWN & $3.97 \times10^{-9}$ & 1.55 & 750 \\ \hline
        MSP & $1.97 \times 10^{-9}$ & 1.50 & 800 \\ \hline
    \end{tabular}
    \caption{Best-fit 'extra component' injection parameters; see Eq.~\ref{eq:inj_extra} for details. $Q_0$ is given at 1 GeV.}
    \label{tab:extra}
\end{table}

\begin{figure}
    \centering
    \includegraphics[width=0.6\textwidth]{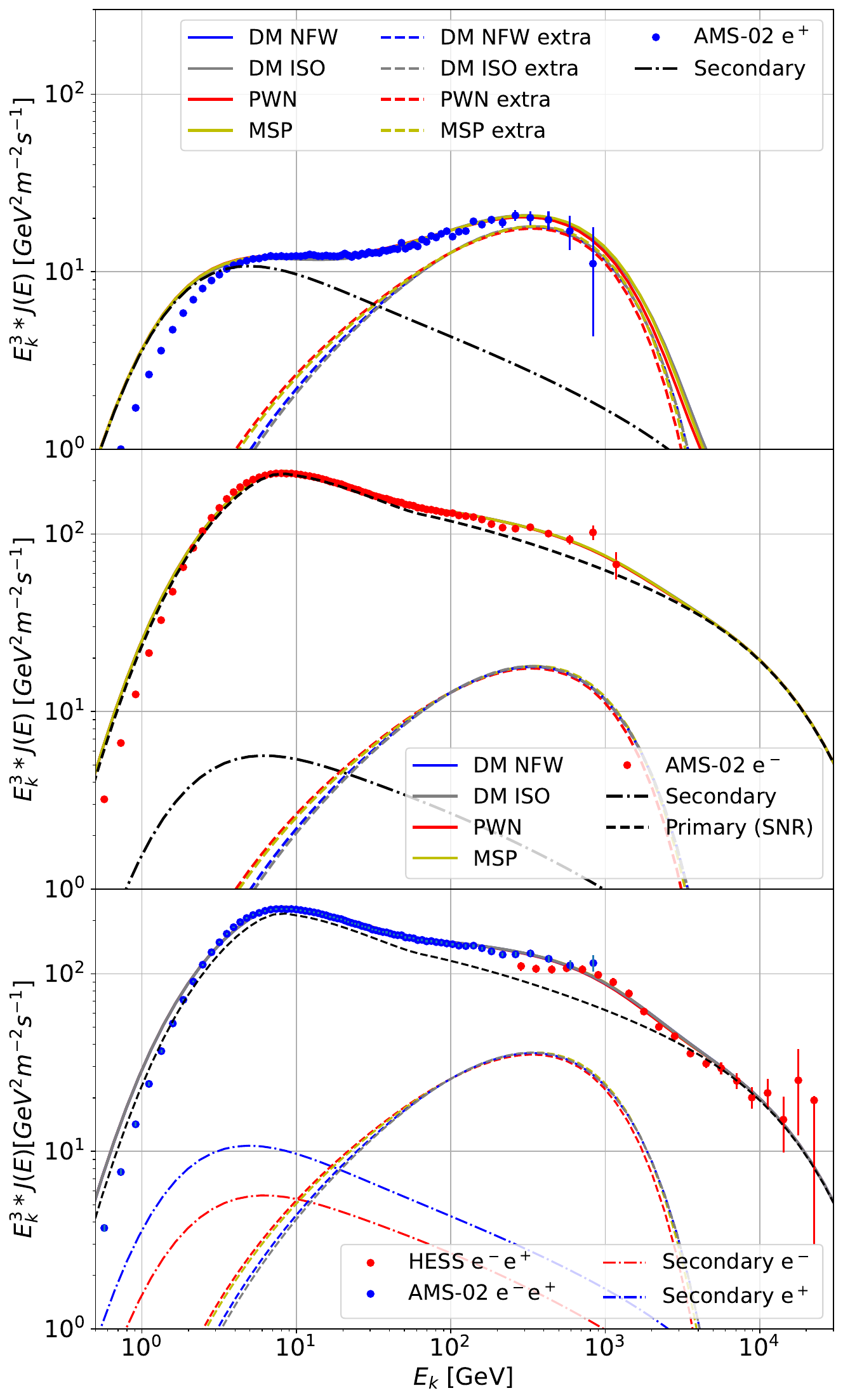}
    \caption{CR positron (top), electron (middle) and all-lepton (bottom) spectra including our 'Extra' component for the three different scenarios. The dashed coloured lines show the extra component. For e$^-$, the black dashed line shows the baseline primary component, while for both e$^-$ and e$^+$, the black dash-dotted line shows the baseline secondary component.}
    \label{fig:extra}
\end{figure}

Regarding the extra component, we look for the normalisation, injection index and cut-off energy that result in the best fit of the positron and electron spectra observed at the Solar neighbourhood for each one of the adopted source morphologies (PWN, MSP, DM NFW and DM ISO).
The numerical results are given in Table~\ref{tab:extra}, and Figure~\ref{fig:extra} shows the comparison between the model predictions and the experimental data.

The injection index and cut-off in the young pulsar scenario are 1.55 and 750 GeV, respectively, which is in good agreement with the expected PWNe index injection from gamma-ray observations, between $1.2-2.2$ \cite{DiMauro_2019,Xi_2019}. 
In the case of the MSP, our result of 1.5 also fits well in the typical range ($1.5-2.5$) used for MSPs \citep{Petrovic_2015, Macias_2021}. 
The two considered DM scenarios are consistent with $\alpha = -1.30$ and $E_{\rm cut} = 900$~GeV. Although there is a large difference between both DM distributions in the innermost region of the Galaxy, at larger Galactic radii their behaviour is quite similar. Therefore, both models yield the same set of injection parameters.
These values are more difficult to interpret, since we do not assume a theoretical annihilation spectrum, but the cut-off energy hints towards a DM particle mass of the order of $\sim 1-10$~TeV.

\begin{figure}
    \centering
    \begin{subfigure}{0.49\textwidth}
        \includegraphics[width = \textwidth]{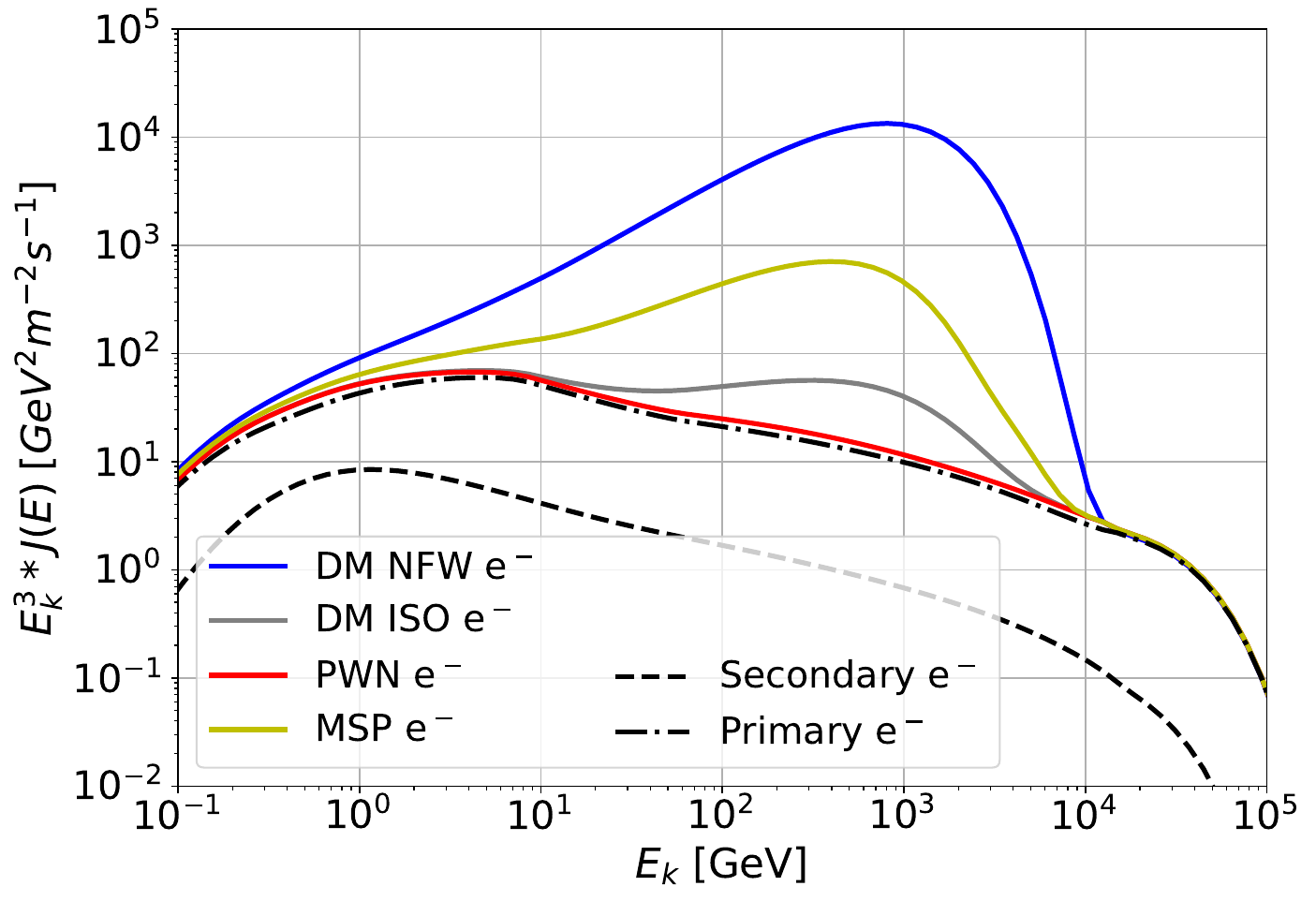}
    \end{subfigure}
    \begin{subfigure}{0.49\textwidth}
        \includegraphics[width =\textwidth]{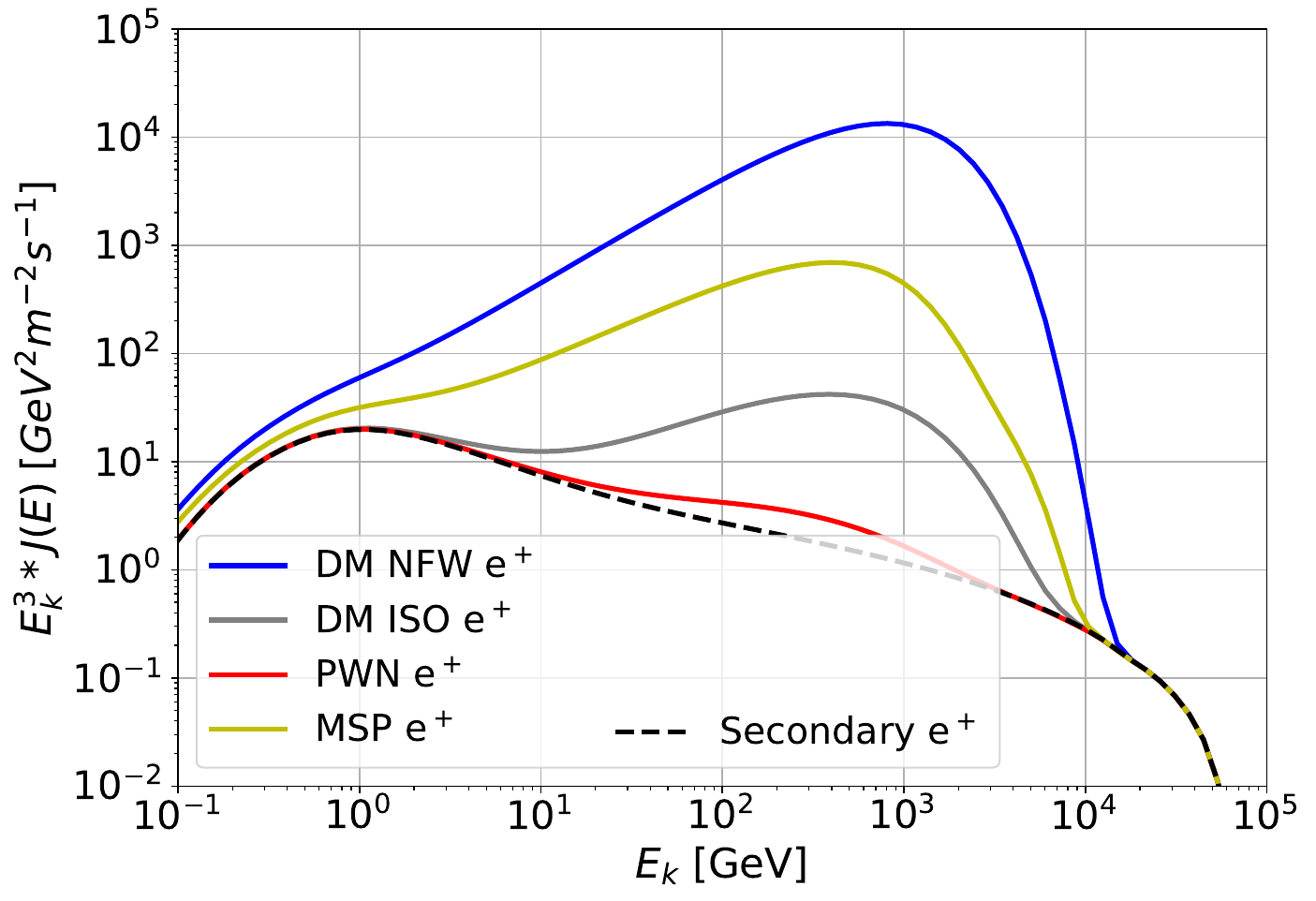}
    \end{subfigure}
    \caption{CR electron (left) and positron (right) spectra at the Galactic center. Solid lines represent the sum of every component in the model.}
    \label{fig:lep_center}
\end{figure}

By construction, all the extra component models yield similar results in the Solar vicinity.
However, large differences are expected both at the Galactic centre, as well as far from the Galactic plane, due to their different spatial distribution.
Figure~\ref{fig:lep_center} shows the resulting CR electron (left) and positron (right) spectra at the center of the Milky Way.
The effect of introducing the extra $e^-e^+$ pairs according to our four prescriptions is clearly noticeable.
While the MSP and both DM scenarios produce a significant boost in the electron and positron content over a broad energy range, PWN represent a minor correction with respect to the baseline model accounting for SNR injection alone.

%-- GAMMA RAYS

\section{Gamma rays}
\label{sec:gamma}

Having obtained a description of the distribution of CRs within the Galaxy, we now proceed to compute their associated gamma-ray emission.
To do this, we make use of the HERMES\footnote{\url{https://github.com/cosmicrays/hermes}} code \cite{HERMES}, where we include different processes: bremsstrahlung and Inverse Compton Scattering (ICS), coming from leptons, and $\pi^0$ decay, from protons and helium.
In addition, they all include the extragalactic gamma-ray background (EGRB) described in \cite{Ackermann_2015, IGRB}.

The \textit{Fermi}-LAT data is extracted using an observation time ranging from 04-08-2008 to 31-12-2020, selecting CLEAN events from the PASS8 data and the P8R3$\_$CLEAN$\_$V3 version of the instrument response function. Events with zenith angle $\theta_z$ $<$ 100$^\circ$ and events given when the LAT instrument was at rocking angles $\theta_r$ $<$ 52$^\circ$ are removed. Our resulting skymaps are calculated using a Healpix projection parameter $nside$ = 256, in order to approximately match the LAT data resolution at high energies.

Since all the scenarios share the same baseline hadronic component, they feature the same $\pi^0$ results, which can be considered as a minimal baseline model. Our CO-to-H2 conversion factor $X_{CO}$ is lower than the values used in recent studies \cite[e.g.][]{Pohl_2008, DeLaTorre_2023}, where even a variable factor depending on the Galactic radius is considered. The gamma-ray emission of the baseline SNR model can be seen in Appendix~\ref{app:snr_gamma}.

\begin{figure}
    \centering
    \begin{subfigure}{0.49\textwidth}
        \includegraphics[width = \textwidth]{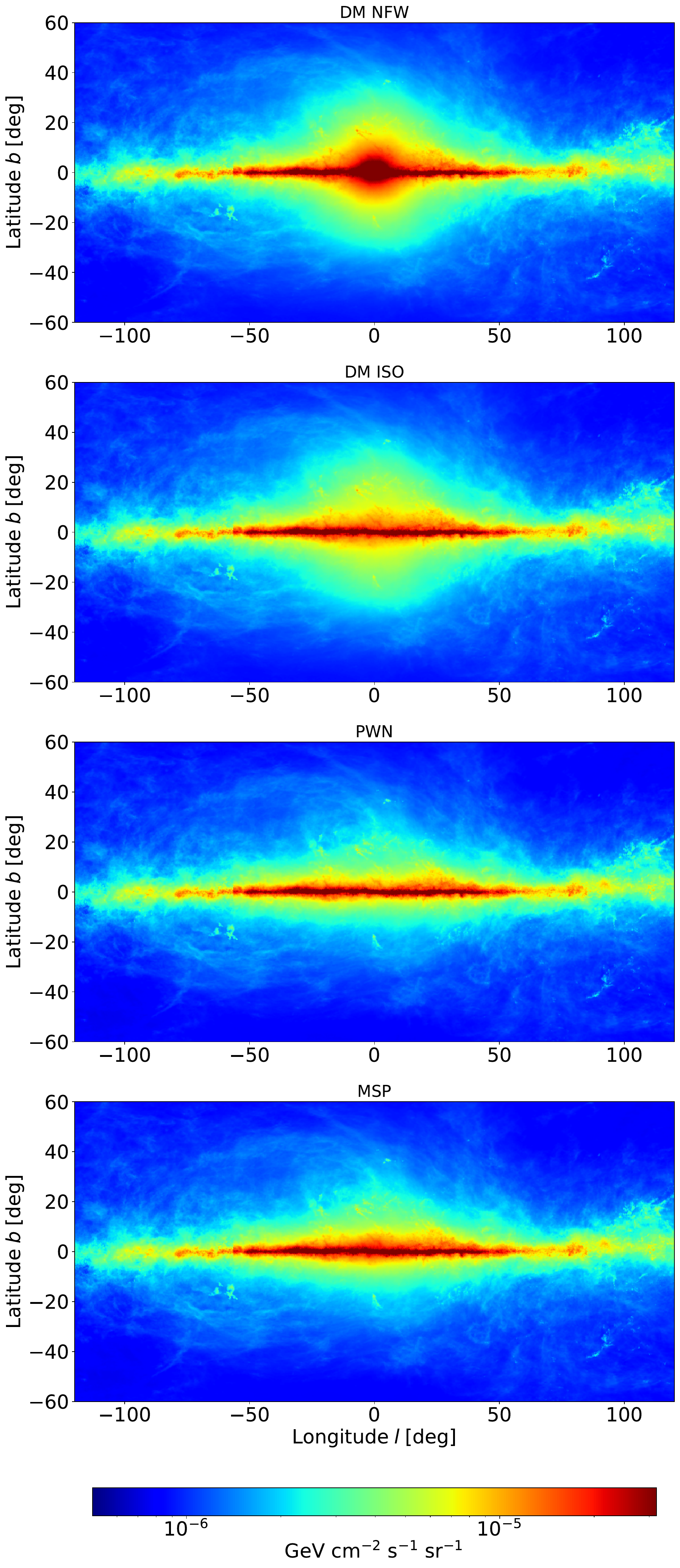}
    \end{subfigure}
    \begin{subfigure}{0.49\textwidth}
        \includegraphics[width = \textwidth]{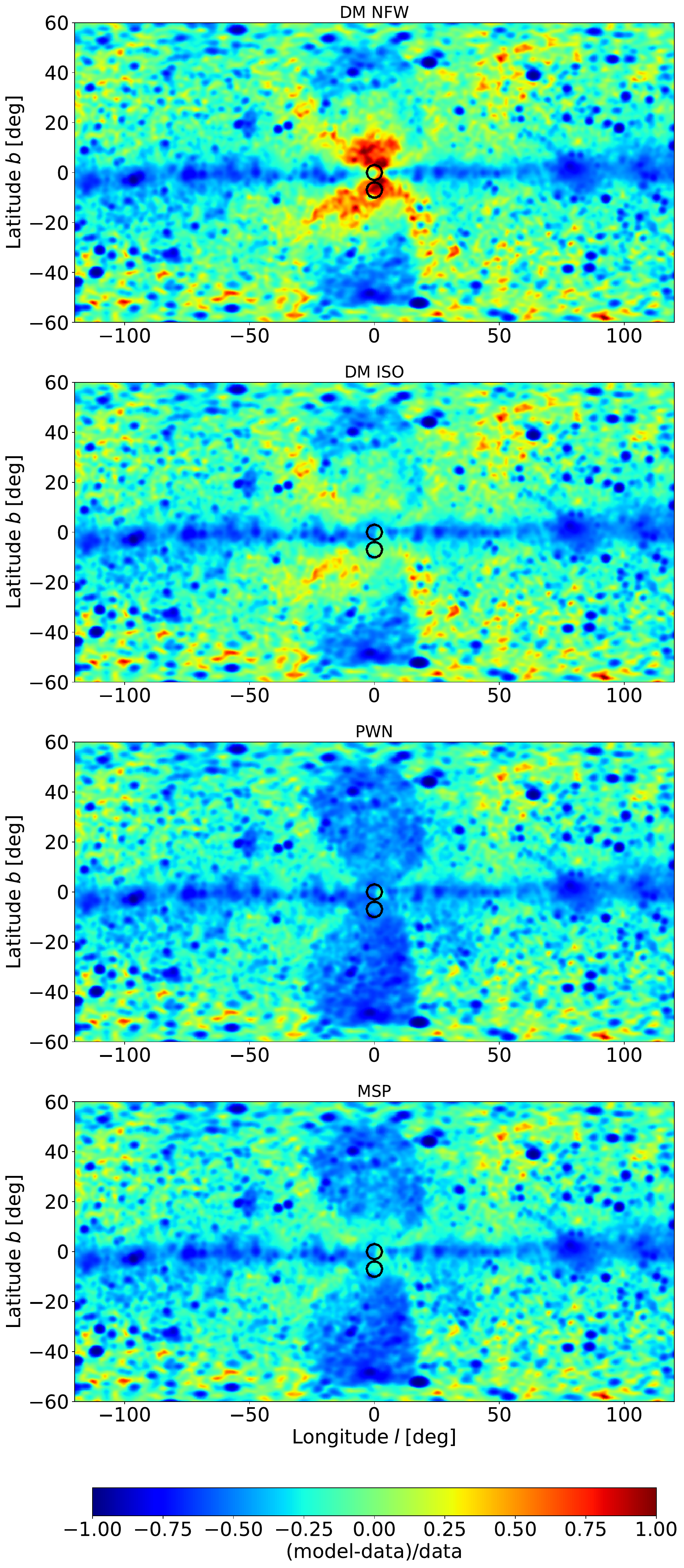}
    \end{subfigure}
    \caption{On the left column, total gamma-ray emission at $10-100$ GeV predicted by our models (from top to bottom: DM NFW, DM ISO, PWN, and MSP). The right column shows the residuals $(model-data)/data$ with respect to the Fermi-LAT data. Black circles represent the regions where we study the gamma-ray spectra (see Figure~\ref{fig:gamma_spec}).}
    \label{fig:resid}
\end{figure}

We show our gamma-ray maps and their residuals with respect to \textit{Fermi}-LAT data in the energy range 10-100 GeV in the left and right columns of Figure~\ref{fig:resid}, respectively.
We perform a 2$^\circ$ Gaussian smoothing in order to mitigate random statistical fluctuations.
One may see the effect of point sources as compact blue circles in the residual images, but modelling these components is beyond the scope of this work.

Away from point sources, all our scenarios are able to reproduce the observations with an accuracy of the order of $\sim 30\%$, with a few exceptions.
On the one hand, the PWN and MSP scenarios clearly need the introduction of another physical component (the so-called 'Fermi bubbles' \cite{Su_2010,Ackermann_2014}) in order to match the data above and below the center of the Milky Way.
On the other hand, DM annihilation within a cuspy (NFW) halo overshoots the observed emission by a factor of $\sim 2$ at intermediate latitudes ($|b| \sim 10^\circ$).
This problem may be mitigated by invoking a cored, 'DM ISO' halo.
Note that at higher latitudes ($|b| \lesssim 50^\circ$), DM annihilation would be able to account for a large fraction of the gamma-ray emission that is usually attributed to the Fermi bubbles.

\begin{sidewaysfigure}
%\begin{figure}
    \centering
    \includegraphics[width = \textwidth]{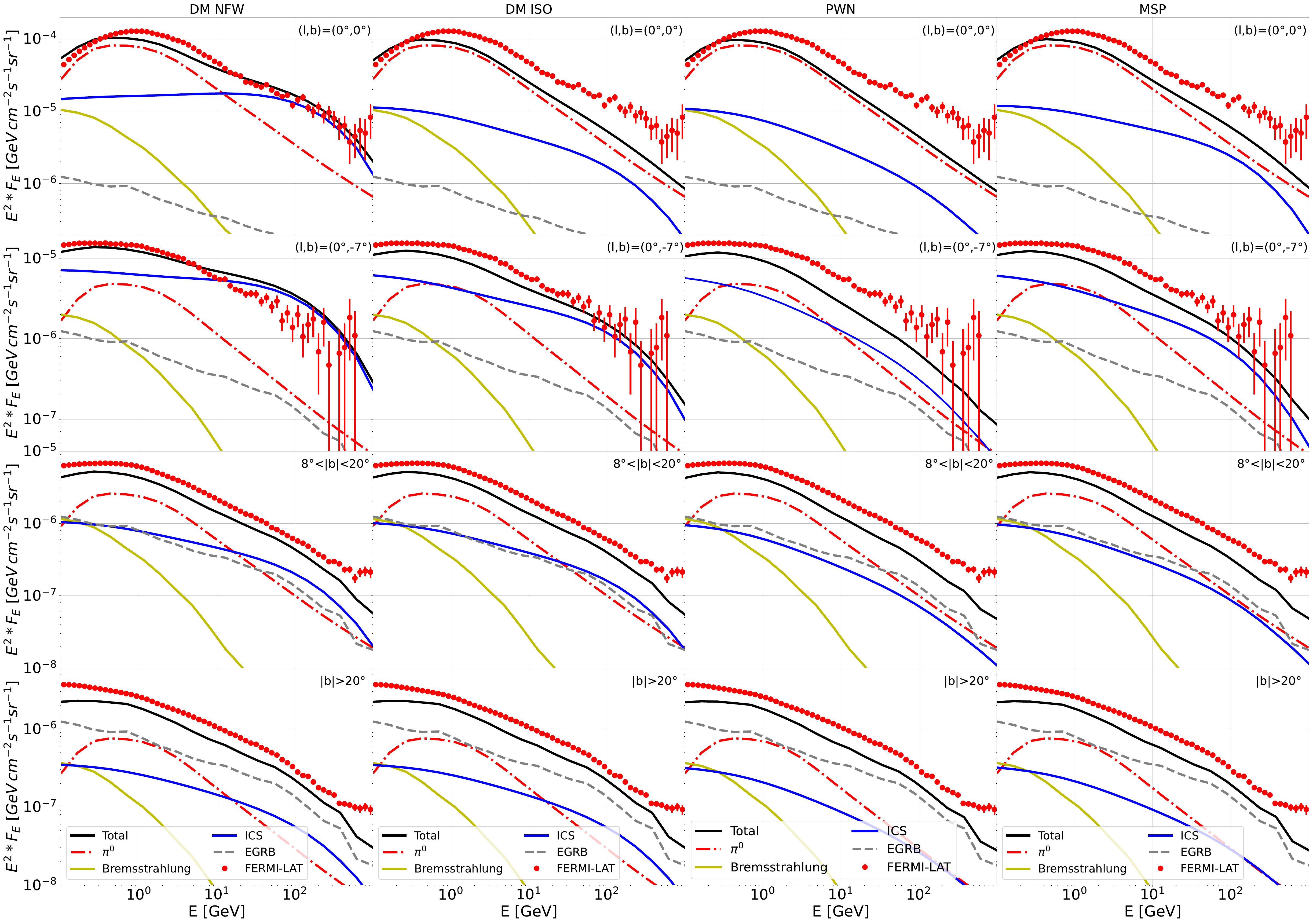}
    \caption{Gamma-ray spectra for every model (from left to right: DM NFW, DM ISO, PWN, MSP) at different sky regions (from top to bottom: Galactic center, $l = 0^\circ$ and $b = -7^\circ$, $8^\circ < |b| <20^\circ$, $|b|>20^\circ$). Data points are from Fermi LAT.}
    \label{fig:gamma_spec}
%\end{figure}
\end{sidewaysfigure}

We also study the spectra of our models at different regions of the sky, which are shown in Figure~\ref{fig:gamma_spec}.
Although our approach does not aim to accurately reproduce the observational measurements, it is able to unambiguously highlight significant discrepancies.
At the center of the Galaxy (top row), the DM NFW model reproduces well the spectra, while the others are clearly below the data above 1 GeV.
On the other hand, the excess of the DM NFW scenario at energies $\gtrsim 3$~GeV is evident, e.g., at $l=0^\circ$, $b=-7^\circ$ (second row).
Since our analysis does not include the prompt emission of gamma-rays, which may actually dominate over the ICS component for DM annihilation, a cuspy halo would be in tension with the gamma-ray observations if the annihilation of DM particles was responsible for a significant fraction of the high-energy positrons.
A more detailed analysis, including this component, could be carried out in order to constrain the inner slope of the density profile. Based on our study, a cored DM profile such as our DM ISO model cannot be confidently ruled out.
On the other hand, both PWN and MSP pulsar scenarios are perfectly compatible with the data, yet in this case one must necessarily invoke additional sources to explain the gamma-ray sky.

At variance with previous studies \cite{Ackermann_2012, Thaler_2023}, where the $\pi^0$ component dominates at every region of the Galaxy, one can readily see in Figure~\ref{fig:gamma_spec} that ICS from the extra source of positrons may make a sizable contribution to the observed gamma-ray emission. For instance, at $l=0^\circ$, $b=-7^\circ$ (second row) and $E = 100$~GeV, it is actually the dominant component for all the scenarios that we have considered.

%-- DISCUSSION --
\section{Discussion}
\label{sec:discussion}

In principle, any of the proposed scenarios could be, at least, partially responsible for the local positron excess.
Although the excess of the NFW scenario with respect to the observed emission is relatively mild (taking into account the uncertainties in the model), the missing gamma-ray components make it significant. In this regard, a cored DM (ISO) halo seems more compatible with the data.
Prompt emission of gamma rays should be investigated in order to derive more stringent constraints on the allowed density profiles, as already done in \cite{Tavakoli_2014, Boudaud_2015} when finding non compatibility with the DM hypothesis. 
Given the uncertainties in the astrophysical ingredients of our model (e.g. gas density, radiation and magnetic fields across the Milky Way), we also consider that a more thorough exploration of the parameter space is necessary if we want to completely and confidently rule out the DM annihilation hypothesis~\cite{cirelli2015status}.

On the opposite extreme, PWN are the only explanation of the local positron excess that would not have a significant gamma-ray imprint.
While this is a positive trait in terms of compatibility, it implies that a more complicated model, with a combination of sources, would be needed to reproduce the observed gamma-ray emission. 
Thus, using the Occam's razor argument, PWN would be disfavoured if another scenario could be found that self-consistently accounted for both the local spectra and the Galactic gamma ray sky.
Since our approach is quite model independent, this result applies to any explanation of the local positron excess associated to the young stellar population or simply concentrated near the Galactic plane for whatever reason.
In this sense, MSP, or any other candidate source that follows the distribution of the old stellar population, provides a much better match in the central regions, and therefore these types of scenarios are certainly worth further scrutiny.

In addition to these considerations, derived from the predicted gamma-ray emission, one may take our phenomenological approach one step further and assess whether the best-fitting parameters that we adopted are realistic, or at least compatible with our current knowledge.
Even if the injection spectrum is tuned to match the local measurements, additional constraints may be obtained based on the energy budget available to each scenario.

To calculate the energy input, we integrate the source spectrum over the whole Galaxy. In our PWN model, the total power injected across the whole Galaxy is $(dE/dt)_{\rm PWN} = 7.2 \times 10^{37}$~erg/s.
There are thousands of known young pulsars in the Milky Way, many of which are expected to host PWN and/or gamma-ray halos (see e.g. \cite{Martin_22}).
The average power per single source is thus of the order of $10^{34}-10^{35}$~erg/s, consistent with observed spin-down luminosities.
For the MSP scenario, the total injected power increases to $(dE/dt)_{\rm MSP} = 1.5 \times 10^{38}$~erg/s.
Although only $\sim 50$ MSP have been discovered so far, 
it is estimated that their total number could be of the order of $(2-7) \times 10^3$ \cite{Linares_2021}, yielding a similar average power per individual source.
Whether this number is realistic or not depends on the physical mechanism of pair production.
Our results are consistent with \cite{Manconi_2020}, who advocate for a 10 per cent efficiency in the conversion of spin-down luminosity in PWN.
For MSP, somewhat lower values ($1-10\%$) are reported by \cite{Linares_2021}. In such a case, the gamma-ray emission could be significantly smaller than predicted by our model.

For DM annihilation, the injected power in form of $e^-e^+$ pairs needed to reproduce the local positron excess amounts to $(dE/dt)_{\rm NFW} = 6.9\times 10^{38}$~erg/s and $(dE/dt)_{\rm ISO} = 4.1\times 10^{38}$ erg/s.
In the Solar neighbourhood, both density profiles inject a power density per unit volume $(du/dt)_{\rm DM,\odot} = 6.3\times 10^{-28}$ GeV/cm$^3$/s. This power density results from integrating in energy Eq.~\ref{eq:inj_extra} at the Solar position and is to be compared with the theoretical rate
\begin{equation}
 \left(\frac{du}{dt}\right)_{\rm DM,\odot} \approx\ \frac{\rho_{\rm DM,\odot}^2}{m_{\rm DM}} \langle\sigma v\rangle c^2.
\end{equation}
Assuming a DM particle mass of 1 TeV (as suggested by our fits) and a local DM density $\rho_{\rm DM,\odot}\, c^2 = 0.4$~GeV~cm$^{-3}$ \cite{Catena_2010}, we have $\langle\sigma v\rangle \approx 3.9\times 10^{-24}$~cm$^3$/s for both models, which is two orders of magnitude above the thermal annihilation cross-section, consistent with the constraints obtained by \cite{WechakamaAscasibar11} for the injection of electron-positron pairs with identical initial energy $E_0 = m_{\rm DM} c^2$ and \cite{WechakamaAscasibar14} for different annihilation channels.

This value may be regarded as a lower limit if we take into account that we are considering only the injected leptons, while part of the energy in a DM annihilation is actually emitted as prompt gamma-ray photons. Even in this case, the result is already in conflict with the current constraints for a cuspy DM distribution \cite{constraints_Hess}.
On the other hand, let us note that a particle with $m_{\rm DM}\, c^2 \sim 1-100$~GeV and $\langle\sigma v\rangle \sim 3\times 10^{-26}$~cm$^3$/s, with a non-negligible leptonic branching ratio (after accounting for prompt emission and other annihilation channels), might be responsible for a fraction of both the positrons in the Solar neighbourhood \cite{Krommydas_2023} and the Galactic gamma-rays in this energy range.

% -- CONCLUSION --

\section{Conclusions}
\label{sec:conclusions}

We have studied the positron excess in the local CR spectrum~\citep{PAMELA}, whose origin might be related to new physics or may unveil new astrophysical mechanisms at play in the Milky Way.
In addition to the secondary positron population produced in CR collisions with the Galactic gas, in this proof-of-concept work we introduced an extra source that injects symmetrically $e^+e^-$ pairs. We looked for the parameters that best reproduced the local measurements of the positron excess by AMS-02 \cite{AMS_positrons} in a phenomenological and model-independent way.
We considered four different scenarios for the nature of the positrons source and its distribution within the Galaxy: two for annihilation of DM particles, with squared NFW \citep{NFW} and isothermal \cite{Begeman_1991} DM halo density profiles, and two related to the pulsar population; PWN following a Lorimer \citep{Lorimer} profile, and MSP (assumed to trace old stars) with a McMillan \citep{McMillan} profile. As they were all fitted to the local spectra, they yield the same positron spectrum at the Earth position, but not in other places of the Galaxy.
Then, we computed the gamma-ray emission associated to each scenario and compared the predicted all-sky maps with the available data from \textit{Fermi}-LAT, to test whether the different scenarios proposed could explain the positron emission observed at Earth without being in conflict with gamma-ray data in other Galactic regions.

We first built a SNR model from fits to CR data. Then, we added to this model the emission of ICS and bremsstrahlung coming from the electrons injected by SNRs and the extra source fitting AMS-02 positron data. These results should be taken as a minimal model, as not every possible contribution is included. The $X_{CO}$ value that we used for the molecular gas density is rather low, leading to a low $\pi^0$ contribution. Neither the gamma-ray point-like sources, the Fermi Bubbles nor DM prompt emission were included. Therefore, any overshoot of the data means that there are significant tensions, even with the high uncertainties in the modelling.

Regarding the gamma-ray results, the predicted emission from our DM NFW scenario fits well the data in the center of the Galaxy. However, it has a significant tension at intermediate latitudes, where the DM ISO scenario has a better agreement with the data. In the case of PWN and MSP, they are both below the data in both regions of the Galaxy. This means that there is some freedom for the addition of the components that are missing in our minimal model. 
The energetics of the PWN and MSP scenarios are consistent with the observed luminosities of single sources. In the case of DM, the resulting cross-section of 3.9$\times$10$^{-24}$ cm$^3$/s is higher than the allowed values by the current constraints \cite{constraints_Hess} for a cuspy, NFW-like DM distribution. Although the constraints for a cored ISO profile should be weaker, there might also be some tension with our result, since we are not considering the prompt DM emission, i.e. the usually dominant component in the gamma-ray domain.
Besides that, our results might be useful to set additional constraints even in the models where the prompt emission is suppressed \cite[e.g.][]{Barak_2023}.

None of our single scenarios is able to simultaneously fit the positron excess and the Galactic center emission at the same time. However, with this proof-of-concept work and without exploring the parameter space in a sophisticated way, most of our results are compatible with the data. 
We cannot rule out any of the scenarios given all the uncertainties in the modelling, but the data at intermediate latitudes certainly shows a preference for a cored DM distribution over a cuspy one. Furthermore, and regardless of the actual physical origin of the extra positrons, we found that the contribution from ICS emission is expected to be significant at intermediate latitudes.

All in all, and given all these findings, we are optimistic about our aim of connecting and explaining simultaneously both the positron excess and the whole gamma-ray sky. Extending the energy range and considering other wavelengths will be very helpful for this purpose. Besides this, a more realistic setup combining our studied scenarios might yield more satisfactory results. These possibilities, together with the addition of the missing gamma-ray components, will be explored out in a future work.

% -------------------------------------------------------------------------
\section*{Acknowledgements}
% -------------------------------------------------------------------------

MR dedicates this work to the memory of his father, Manuel Rocamora Lozano. YA acknowledges financial support from grant PID2019-107408GB-C42 of the Spanish State Research Agency (AEI/10.13039/501100011033). A significant fraction of this work has been carried out during a research visit to the International Centre for Radio Astronomy Research -- University of Western Australia (ICRAR-UWA) funded from the mobility programme \emph{Ayudas de recualificación del profesorado universitario funionario o contratado} (CA2/RSUE/2021-00817) of the \emph{Universidad Autónoma de Madrid} (UAM), within the framework of the \emph{Plan de Recuperación, Transformación y Resiliencia} of the \emph{Ministerio de Universidades} (Spain). The work of MASC was supported by the grants PID2021-125331NB-I00 and CEX2020-001007-S, both funded by MCIN/AEI/10.13039/501100011033 and by ``ERDF A way of making Europe’'. MASC also acknowledges the MultiDark Network, ref. RED2022-134411-T. The work of MW was supported by Kasetsart University Research and Development Institute (KURDI). During the first stages of this project PDL was supported by the Swedish Research Council under contract 2019-05135. PDL is currently supported by the Juan de la Cierva JDC2022-048916-I grant.

\bibliographystyle{unsrt}%{mnras}
\bibliography{bibliography.bib}

\begin{thebibliography}{100}

\bibitem{Blasi_2013}
Pasquale {Blasi}.
\newblock {The origin of galactic cosmic rays}.
\newblock {\em \aapr}, 21:70, November 2013.

\bibitem{Moskalenko_1998}
I.~V. {Moskalenko} and A.~W. {Strong}.
\newblock {Production and Propagation of Cosmic-Ray Positrons and Electrons}.
\newblock {\em \apj}, 493(2):694--707, January 1998.

\bibitem{Prantzos_2011}
N.~{Prantzos}, C.~{Boehm}, A.~M. {Bykov}, R.~{Diehl}, K.~{Ferri{\`e}re},
  N.~{Guessoum}, P.~{Jean}, J.~{Knoedlseder}, A.~{Marcowith}, I.~V.
  {Moskalenko}, A.~{Strong}, and G.~{Weidenspointner}.
\newblock {The 511 keV emission from positron annihilation in the Galaxy}.
\newblock {\em Reviews of Modern Physics}, 83(3):1001--1056, July 2011.

\bibitem{Vincent_2012}
Aaron~C Vincent, Pierrick Martin, and James~M Cline.
\newblock Interacting dark matter contribution to the galactic 511 kev gamma
  ray emission: constraining the morphology with integral/spi observations.
\newblock {\em Journal of Cosmology and Astroparticle Physics},
  2012(04):022–022, April 2012.

\bibitem{luque2023multimessenger}
Pedro~De la~Torre~Luque, Shyam Balaji, and Pierluca Carenza.
\newblock Multimessenger search for electrophilic feebly interacting particles
  from supernovae, 2023.

\bibitem{PAMELA}
O.~{Adriani} et~al.
\newblock {An anomalous positron abundance in cosmic rays with energies
  1.5-100GeV}.
\newblock {\em \nat}, 458(7238):607--609, April 2009.

\bibitem{AMS_positrons}
M.~{Aguilar} et~al.
\newblock {Electron and Positron Fluxes in Primary Cosmic Rays Measured with
  the Alpha Magnetic Spectrometer on the International Space Station}.
\newblock {\em \prl}, 113(12):121102, September 2014.

\bibitem{FERMI_positrons}
S.~{Abdollahi} et~al.
\newblock {Cosmic-ray electron-positron spectrum from 7 GeV to 2 TeV with the
  Fermi Large Area Telescope}.
\newblock {\em \prd}, 95(8):082007, April 2017.

\bibitem{Diesing_2020}
Rebecca {Diesing} and Damiano {Caprioli}.
\newblock {Nonsecondary origin of cosmic ray positrons}.
\newblock {\em \prd}, 101(10):103030, May 2020.

\bibitem{Hooper_2009}
Dan {Hooper}, Pasquale {Blasi}, and Pasquale~Dario {Serpico}.
\newblock {Pulsars as the sources of high energy cosmic ray positrons}.
\newblock {\em \jcap}, 2009(1):025, January 2009.

\bibitem{Hooper_2017}
Dan {Hooper}, Ilias {Cholis}, Tim {Linden}, and Ke~{Fang}.
\newblock {HAWC observations strongly favor pulsar interpretations of the
  cosmic-ray positron excess}.
\newblock {\em \prd}, 96(10):103013, November 2017.

\bibitem{Cholis_2018}
Ilias {Cholis}, Tanvi {Karwal}, and Marc {Kamionkowski}.
\newblock {Features in the spectrum of cosmic-ray positrons from pulsars}.
\newblock {\em \prd}, 97(12):123011, June 2018.

\bibitem{DiMauro_2019}
Mattia {Di Mauro}, Silvia {Manconi}, and Fiorenza {Donato}.
\newblock {Detection of a {\ensuremath{\gamma}} -ray halo around Geminga with
  the Fermi-LAT data and implications for the positron flux}.
\newblock {\em \prd}, 100(12):123015, December 2019.

\bibitem{Manconi_2020}
Silvia {Manconi}, Mattia {Di Mauro}, and Fiorenza {Donato}.
\newblock {Contribution of pulsars to cosmic-ray positrons in light of recent
  observation of inverse-Compton halos}.
\newblock {\em \prd}, 102(2):023015, July 2020.

\bibitem{Fornieri_2020}
O.~{Fornieri}, D.~{Gaggero}, and D.~{Grasso}.
\newblock {Features in cosmic-ray lepton data unveil the properties of nearby
  cosmic accelerators}.
\newblock {\em \jcap}, 2020(2):009, February 2020.

\bibitem{Evoli_2021}
Carmelo {Evoli}, Elena {Amato}, Pasquale {Blasi}, and Roberto {Aloisio}.
\newblock {Galactic factories of cosmic-ray electrons and positrons}.
\newblock {\em \prd}, 103(8):083010, April 2021.

\bibitem{Boudaud_2015}
M.~{Boudaud}, S.~{Aupetit}, S.~{Caroff}, A.~{Putze}, G.~{Belanger},
  Y.~{Genolini}, C.~{Goy}, V.~{Poireau}, V.~{Poulin}, S.~{Rosier}, P.~{Salati},
  L.~{Tao}, and M.~{Vecchi}.
\newblock {A new look at the cosmic ray positron fraction}.
\newblock {\em \aap}, 575:A67, March 2015.

\bibitem{Jin_2015}
Hong-Bo {Jin}, Yue-Liang {Wu}, and Yu-Feng {Zhou}.
\newblock {Cosmic ray propagation and dark matter in light of the latest AMS-02
  data}.
\newblock {\em \jcap}, 2015(9):049--049, September 2015.

\bibitem{Jin_2020}
Hong-Bo {Jin}, Yue-Liang {Wu}, and Yu-Feng {Zhou}.
\newblock {Astrophysical Background and Dark Matter Implication Based on Latest
  AMS-02 Data}.
\newblock {\em \apj}, 901(1):80, September 2020.

\bibitem{Zhan_2023}
Haoxiang {Zhan}.
\newblock {Constraining the dark matter interpretation of the positron excess
  with $\gamma$-ray data}.
\newblock {\em arXiv e-prints}, page arXiv:2305.01992, May 2023.

\bibitem{Goodenough_2009}
Lisa {Goodenough} and Dan {Hooper}.
\newblock {Possible Evidence For Dark Matter Annihilation In The Inner Milky
  Way From The Fermi Gamma Ray Space Telescope}.
\newblock {\em arXiv e-prints}, page arXiv:0910.2998, October 2009.

\bibitem{Hooper_2011a}
Dan {Hooper} and Lisa {Goodenough}.
\newblock {Dark matter annihilation in the Galactic Center as seen by the Fermi
  Gamma Ray Space Telescope}.
\newblock {\em Physics Letters B}, 697(5):412--428, March 2011.

\bibitem{Hooper_2011b}
Dan {Hooper} and Tim {Linden}.
\newblock {Origin of the gamma rays from the Galactic Center}.
\newblock {\em \prd}, 84(12):123005, December 2011.

\bibitem{Boyarsky_2011}
Alexey {Boyarsky}, Denys {Malyshev}, and Oleg {Ruchayskiy}.
\newblock {A comment on the emission from the Galactic Center as seen by the
  Fermi telescope}.
\newblock {\em Physics Letters B}, 705(3):165--169, November 2011.

\bibitem{Abazajian_2012}
Kevork~N. {Abazajian} and Manoj {Kaplinghat}.
\newblock {Detection of a gamma-ray source in the Galactic Center consistent
  with extended emission from dark matter annihilation and concentrated
  astrophysical emission}.
\newblock {\em \prd}, 86(8):083511, October 2012.

\bibitem{Gordon_2013}
Chris {Gordon} and Oscar {Mac{\'\i}as}.
\newblock {Dark matter and pulsar model constraints from Galactic Center
  Fermi-LAT gamma-ray observations}.
\newblock {\em \prd}, 88(8):083521, October 2013.

\bibitem{Abazajian_2014}
Kevork~N. {Abazajian}, Nicolas {Canac}, Shunsaku {Horiuchi}, and Manoj
  {Kaplinghat}.
\newblock {Astrophysical and dark matter interpretations of extended gamma-ray
  emission from the Galactic Center}.
\newblock {\em \prd}, 90(2):023526, July 2014.

\bibitem{Calore_2015a}
Francesca {Calore}, Ilias {Cholis}, Christopher {McCabe}, and Christoph
  {Weniger}.
\newblock {A tale of tails: Dark matter interpretations of the Fermi GeV excess
  in light of background model systematics}.
\newblock {\em \prd}, 91(6):063003, March 2015.

\bibitem{Calore_2015b}
Francesca {Calore}, Ilias {Cholis}, and Christoph {Weniger}.
\newblock {Background model systematics for the Fermi GeV excess}.
\newblock {\em \jcap}, 2015(3):038--038, March 2015.

\bibitem{Daylan_2016}
Tansu {Daylan}, Douglas~P. {Finkbeiner}, Dan {Hooper}, Tim {Linden}, Stephen
  K.~N. {Portillo}, Nicholas~L. {Rodd}, and Tracy~R. {Slatyer}.
\newblock {The characterization of the gamma-ray signal from the central Milky
  Way: A case for annihilating dark matter}.
\newblock {\em Physics of the Dark Universe}, 12:1--23, June 2016.

\bibitem{Ajello_2016}
M.~{Ajello} et~al.
\newblock {Fermi-LAT Observations of High-Energy Gamma-Ray Emission toward the
  Galactic Center}.
\newblock {\em \apj}, 819(1):44, March 2016.

\bibitem{Ackermann_2017}
M.~{Ackermann} et~al.
\newblock {The Fermi Galactic Center GeV Excess and Implications for Dark
  Matter}.
\newblock {\em \apj}, 840(1):43, May 2017.

\bibitem{DiMauro_2019_gev}
Mattia {Di Mauro}, Xian {Hou}, Christopher {Eckner}, Gabrijela {Zaharijas}, and
  Eric {Charles}.
\newblock {Search for {\ensuremath{\gamma}} -ray emission from dark matter
  particle interactions from the Andromeda and Triangulum galaxies with the
  Fermi Large Area Telescope}.
\newblock {\em \prd}, 99(12):123027, June 2019.

\bibitem{Bartels_2015}
Richard {Bartels}, Suraj {Krishnamurthy}, and Christoph {Weniger}.
\newblock {Strong Support for the Millisecond Pulsar Origin of the Galactic
  Center GeV Excess}.
\newblock {\em \prl}, 116(5):051102, February 2016.

\bibitem{Bartels_2018}
R.~{Bartels}, F.~{Calore}, E.~{Storm}, and C.~{Weniger}.
\newblock {Galactic binaries can explain the Fermi Galactic centre excess and
  511 keV emission}.
\newblock {\em \mnras}, 480(3):3826--3841, November 2018.

\bibitem{Venter_2015}
C.~{Venter}, A.~{Kopp}, A.~K. {Harding}, P.~L. {Gonthier}, and
  I.~{B{\"u}sching}.
\newblock {Cosmic-ray Positrons from Millisecond Pulsars}.
\newblock {\em \apj}, 807(2):130, July 2015.

\bibitem{Linares_2021}
M.~{Linares} and M.~{Kachelrie{\ss}}.
\newblock {Cosmic ray positrons from compact binary millisecond pulsars}.
\newblock {\em \jcap}, 2021(2):030, February 2021.

\bibitem{Acero_2016}
F.~{Acero}, M.~{Ackermann}, and {et al.} {Ajello}, M.
\newblock {Development of the Model of Galactic Interstellar Emission for
  Standard Point-source Analysis of Fermi Large Area Telescope Data}.
\newblock {\em \apjs}, 223(2):26, April 2016.

\bibitem{Storm_2017}
Emma {Storm}, Christoph {Weniger}, and Francesca {Calore}.
\newblock {SkyFACT: high-dimensional modeling of gamma-ray emission with
  adaptive templates and penalized likelihoods}.
\newblock {\em \jcap}, 2017(8):022, August 2017.

\bibitem{Pohl_2022}
Martin {Pohl}, Oscar {Macias}, Phaedra {Coleman}, and Chris {Gordon}.
\newblock {Assessing the Impact of Hydrogen Absorption on the Characteristics
  of the Galactic Center Excess}.
\newblock {\em \apj}, 929(2):136, April 2022.

\bibitem{Ackermann_2012}
M.~{Ackermann}, M.~{Ajello}, and {et al.} {Atwood}.
\newblock {Fermi-LAT Observations of the Diffuse {\ensuremath{\gamma}}-Ray
  Emission: Implications for Cosmic Rays and the Interstellar Medium}.
\newblock {\em \apj}, 750(1):3, May 2012.

\bibitem{Orlando_2018}
E.~{Orlando}.
\newblock {Imprints of cosmic rays in multifrequency observations of the
  interstellar emission}.
\newblock {\em \mnras}, 475(2):2724--2742, April 2018.

\bibitem{Cholis_2022}
Ilias {Cholis}, Yi-Ming {Zhong}, Samuel~D. {McDermott}, and Joseph~P.
  {Surdutovich}.
\newblock {Return of the templates: Revisiting the Galactic Center excess with
  multimessenger observations}.
\newblock {\em \prd}, 105(10):103023, May 2022.

\bibitem{Thaler_2023}
J.~{Thaler}, R.~{Kissmann}, and O.~{Reimer}.
\newblock {Cosmic-ray propagation under consideration of a spatially resolved
  source distribution}.
\newblock {\em Astroparticle Physics}, 144:102776, January 2023.

\bibitem{DeLaTorre_2023}
P.~{De La Torre Luque}, D.~{Gaggero}, D.~{Grasso}, O.~{Fornieri}, K.~{Egberts},
  C.~{Steppa}, and C.~{Evoli}.
\newblock {Galactic diffuse gamma rays meet the PeV frontier}.
\newblock {\em \aap}, 672:A58, April 2023.

\bibitem{Trotta_2011}
R.~{Trotta}, G.~{J{\'o}hannesson}, I.~V. {Moskalenko}, T.~A. {Porter}, R.~{Ruiz
  de Austri}, and A.~W. {Strong}.
\newblock {Constraints on Cosmic-ray Propagation Models from A Global Bayesian
  Analysis}.
\newblock {\em \apj}, 729(2):106, March 2011.

\bibitem{Korsmeier_2016}
Michael {Korsmeier} and Alessandro {Cuoco}.
\newblock {Galactic cosmic-ray propagation in the light of AMS-02: Analysis of
  protons, helium, and antiprotons}.
\newblock {\em \prd}, 94(12):123019, December 2016.

\bibitem{Evoli_2019}
Carmelo {Evoli}, Roberto {Aloisio}, and Pasquale {Blasi}.
\newblock {Galactic cosmic rays after the AMS-02 observations}.
\newblock {\em \prd}, 99(10):103023, May 2019.

\bibitem{Evoli_2020}
Carmelo {Evoli}, Giovanni {Morlino}, Pasquale {Blasi}, and Roberto {Aloisio}.
\newblock {AMS-02 beryllium data and its implication for cosmic ray transport}.
\newblock {\em \prd}, 101(2):023013, January 2020.

\bibitem{Luque_2021}
P.~De La~Torre Luque, M.~N. Mazziotta, F.~Loparco, F.~Gargano, and D.~Serini.
\newblock {Markov chain Monte Carlo analyses of the flux ratios of B, Be and Li
  with the DRAGON2 code}.
\newblock {\em JCAP}, 07:010, 2021.

\bibitem{Ginzburg_64}
V.~L. {Ginzburg} and S.~I. {Syrovatskii}.
\newblock {\em {The Origin of Cosmic Rays}}.
\newblock 1964.

\bibitem{Berezinskii_90}
V.~S. {Berezinskii}, S.~V. {Bulanov}, V.~A. {Dogiel}, and V.~S. {Ptuskin}.
\newblock {\em {Astrophysics of cosmic rays}}.
\newblock 1990.

\bibitem{DRAGON}
Carmelo {Evoli}, Daniele {Gaggero}, Andrea {Vittino}, Giuseppe {Di Bernardo},
  Mattia {Di Mauro}, Arianna {Ligorini}, Piero {Ullio}, and Dario {Grasso}.
\newblock {Cosmic-ray propagation with DRAGON2: I. numerical solver and
  astrophysical ingredients}.
\newblock {\em \jcap}, 2017(2):015, February 2017.

\bibitem{fornieri_2021}
Ottavio {Fornieri}, Daniele {Gaggero}, Daniel {Guberman}, Loann {Brahimi},
  Pedro De La~Torre {Luque}, and Alexandre {Marcowith}.
\newblock {Diffusive origin for the cosmic-ray spectral hardening reveals
  signatures of a nearby source in the leptons and protons data}.
\newblock {\em \prd}, 104(10):103013, November 2021.

\bibitem{sync_spec}
Giuseppe {Di Bernardo}, Carmelo {Evoli}, Daniele {Gaggero}, Dario {Grasso}, and
  Luca {Maccione}.
\newblock {Cosmic ray electrons, positrons and the synchrotron emission of the
  Galaxy: consistent analysis and implications}.
\newblock {\em \jcap}, 2013(3):036, March 2013.

\bibitem{AMS_comp}
M.~{Aguilar}, L.~{Ali Cavasonza}, G.~{Ambrosi}, and {AMS Collaboration}.
\newblock {Towards Understanding the Origin of Cosmic-Ray Positrons}.
\newblock {\em \prl}, 122(4):041102, February 2019.

\bibitem{Ferriere}
Katia~M. {Ferri{\`e}re}.
\newblock {The interstellar environment of our galaxy}.
\newblock {\em Reviews of Modern Physics}, 73(4):1031--1066, October 2001.

\bibitem{Lorimer}
D.~R. {Lorimer}, A.~J. {Faulkner}, A.~G. {Lyne}, R.~N. {Manchester},
  M.~{Kramer}, M.~A. {McLaughlin}, G.~{Hobbs}, A.~{Possenti}, I.~H. {Stairs},
  F.~{Camilo}, M.~{Burgay}, N.~{D'Amico}, A.~{Corongiu}, and F.~{Crawford}.
\newblock {The Parkes Multibeam Pulsar Survey - VI. Discovery and timing of 142
  pulsars and a Galactic population analysis}.
\newblock {\em \mnras}, 372(2):777--800, October 2006.

\bibitem{McMillan}
Paul~J. {McMillan}.
\newblock {The mass distribution and gravitational potential of the Milky Way}.
\newblock {\em \mnras}, 465(1):76--94, February 2017.

\bibitem{FloresPrimack94}
Ricardo~A. {Flores} and Joel~R. {Primack}.
\newblock {Observational and Theoretical Constraints on Singular Dark Matter
  Halos}.
\newblock {\em \apjl}, 427:L1, May 1994.

\bibitem{Moore94}
Ben {Moore}.
\newblock {Evidence against dissipation-less dark matter from observations of
  galaxy haloes}.
\newblock {\em \nat}, 370(6491):629--631, August 1994.

\bibitem{DiCintio_2014}
Arianna {Di Cintio}, Chris~B. {Brook}, Andrea~V. {Macci{\`o}}, Greg~S.
  {Stinson}, Alexander {Knebe}, Aaron~A. {Dutton}, and James {Wadsley}.
\newblock {The dependence of dark matter profiles on the stellar-to-halo mass
  ratio: a prediction for cusps versus cores}.
\newblock {\em \mnras}, 437(1):415--423, January 2014.

\bibitem{Benito_2019}
Maria {Benito}, Alessandro {Cuoco}, and Fabio {Iocco}.
\newblock {Handling the uncertainties in the Galactic Dark Matter distribution
  for particle Dark Matter searches}.
\newblock {\em \jcap}, 2019(3):033, March 2019.

\bibitem{NFW}
Julio~F. {Navarro}, Carlos~S. {Frenk}, and Simon D.~M. {White}.
\newblock {The Structure of Cold Dark Matter Halos}.
\newblock {\em \apj}, 462:563, May 1996.

\bibitem{Begeman_1991}
K.~G. {Begeman}, A.~H. {Broeils}, and R.~H. {Sanders}.
\newblock {Extended rotation curves of spiral galaxies : dark haloes and
  modified dynamics.}
\newblock {\em \mnras}, 249:523, April 1991.

\bibitem{Pato_2015}
Miguel {Pato}, Fabio {Iocco}, and Gianfranco {Bertone}.
\newblock {Dynamical constraints on the dark matter distribution in the Milky
  Way}.
\newblock {\em \jcap}, 2015(12):001--001, December 2015.

\bibitem{Bertone_2005}
Gianfranco {Bertone}, Dan {Hooper}, and Joseph {Silk}.
\newblock {Particle dark matter: evidence, candidates and constraints}.
\newblock {\em \physrep}, 405(5-6):279--390, January 2005.

\bibitem{Pshirkov_2011}
M.~S. {Pshirkov}, P.~G. {Tinyakov}, P.~P. {Kronberg}, and K.~J. {Newton-McGee}.
\newblock {Deriving the Global Structure of the Galactic Magnetic Field from
  Faraday Rotation Measures of Extragalactic Sources}.
\newblock {\em \apj}, 738(2):192, September 2011.

\bibitem{HERMES}
A.~{Dundovic}, C.~{Evoli}, D.~{Gaggero}, and D.~{Grasso}.
\newblock {Simulating the Galactic multi-messenger emissions with HERMES}.
\newblock {\em \aap}, 653:A18, September 2021.

\bibitem{HI_survey}
{HI4PI Collaboration}, N.~{Ben Bekhti}, L.~{Fl{\"o}er}, R.~{Keller}, J.~{Kerp},
  D.~{Lenz}, B.~{Winkel}, J.~{Bailin}, M.~R. {Calabretta}, L.~{Dedes}, H.~A.
  {Ford}, B.~K. {Gibson}, U.~{Haud}, S.~{Janowiecki}, P.~M.~W. {Kalberla},
  F.~J. {Lockman}, N.~M. {McClure-Griffiths}, T.~{Murphy}, H.~{Nakanishi},
  D.~J. {Pisano}, and L.~{Staveley-Smith}.
\newblock {HI4PI: A full-sky H I survey based on EBHIS and GASS}.
\newblock {\em \aap}, 594:A116, October 2016.

\bibitem{H2_survey}
T.~M. {Dame}, Dap {Hartmann}, and P.~{Thaddeus}.
\newblock {The Milky Way in Molecular Clouds: A New Complete CO Survey}.
\newblock {\em \apj}, 547(2):792--813, February 2001.

\bibitem{Strong_1996}
A.~W. {Strong} and J.~R. {Mattox}.
\newblock {Gradient model analysis of EGRET diffuse Galactic
  {\ensuremath{\gamma}}-ray emission.}
\newblock {\em \aap}, 308:L21--L24, April 1996.

\bibitem{DRAGON_xsec}
Carmelo {Evoli}, Daniele {Gaggero}, Andrea {Vittino}, Mattia {Di Mauro}, Dario
  {Grasso}, and Mario~Nicola {Mazziotta}.
\newblock {Cosmic-ray propagation with DRAGON2: II. Nuclear interactions with
  the interstellar gas}.
\newblock {\em \jcap}, 2018(7):006, July 2018.

\bibitem{Kamae_2006}
Tuneyoshi {Kamae}, Niklas {Karlsson}, Tsunefumi {Mizuno}, Toshinori {Abe}, and
  Tatsumi {Koi}.
\newblock {Parameterization of {\ensuremath{\gamma}}, e$^{+/-}$, and Neutrino
  Spectra Produced by p-p Interaction in Astronomical Environments}.
\newblock {\em \apj}, 647(1):692--708, August 2006.

\bibitem{AMS}
M.~{Aguilar}, D.~{Aisa}, B.~{Alpat}, and {AMS Collaboration}.
\newblock {Precision Measurement of the Proton Flux in Primary Cosmic Rays from
  Rigidity 1 GV to 1.8 TV with the Alpha Magnetic Spectrometer on the
  International Space Station}.
\newblock {\em \prl}, 114(17):171103, May 2015.

\bibitem{Aharonian_2008}
F.~{Aharonian} et~al.
\newblock {Energy Spectrum of Cosmic-Ray Electrons at TeV Energies}.
\newblock {\em \prl}, 101(26):261104, December 2008.

\bibitem{Voyager}
A.~C. {Cummings}, E.~C. {Stone}, B.~C. {Heikkila}, N.~{Lal}, W.~R. {Webber},
  G.~{J{\'o}hannesson}, I.~V. {Moskalenko}, E.~{Orlando}, and T.~A. {Porter}.
\newblock {Galactic Cosmic Rays in the Local Interstellar Medium: Voyager 1
  Observations and Model Results}.
\newblock {\em \apj}, 831(1):18, November 2016.

\bibitem{Usoskin_2005}
Ilya~G. {Usoskin}, Katja {Alanko-Huotari}, Gennady~A. {Kovaltsov}, and Kalevi
  {Mursula}.
\newblock {Heliospheric modulation of cosmic rays: Monthly reconstruction for
  1951-2004}.
\newblock {\em Journal of Geophysical Research (Space Physics)},
  110(A12):A12108, December 2005.

\bibitem{Usoskin_2011}
Ilya~G. {Usoskin}, Galina~A. {Bazilevskaya}, and Gennady~A. {Kovaltsov}.
\newblock {Solar modulation parameter for cosmic rays since 1936 reconstructed
  from ground-based neutron monitors and ionization chambers}.
\newblock {\em Journal of Geophysical Research (Space Physics)},
  116(A2):A02104, February 2011.

\bibitem{DelaTorreLuque:2023zyd}
Pedro De~la Torre~Luque, Mario~Nicola Mazziotta, and Francesco Loparco.
\newblock {The FLUKA cross sections for cosmic-ray leptons and uncertainties on
  current positron predictions}.
\newblock 5 2023.

\bibitem{Koldobskiy_2021}
S.~{Koldobskiy}, M.~{Kachelrie{\ss}}, A.~{Lskavyan}, A.~{Neronov},
  S.~{Ostapchenko}, and D.~V. {Semikoz}.
\newblock {Energy spectra of secondaries in proton-proton interactions}.
\newblock {\em \prd}, 104(12):123027, December 2021.

\bibitem{Orusa_2022}
Luca {Orusa}, Mattia {Di Mauro}, Fiorenza {Donato}, and Michael {Korsmeier}.
\newblock {New determination of the production cross section for secondary
  positrons and electrons in the Galaxy}.
\newblock {\em \prd}, 105(12):123021, June 2022.

\bibitem{Aloisio_2013}
Roberto {Aloisio} and Pasquale {Blasi}.
\newblock {Propagation of galactic cosmic rays in the presence of
  self-generated turbulence}.
\newblock {\em \jcap}, 2013(7):001, July 2013.

\bibitem{Boschini_2020}
M.~J. Boschini, S.~Della Torre, M.~Gervasi, D.~Grandi, G.~JÃ³hannesson, G.~La
  Vacca, N.~Masi, I.~V. Moskalenko, S.~Pensotti, T.~A. Porter, L.~Quadrani,
  P.~G. Rancoita, D.~Rozza, and M.~Tacconi.
\newblock Inference of the local interstellar spectra of cosmic-ray nuclei
  zÂ â‰¤Â 28 with the galpropâ€“helmod framework.
\newblock {\em The Astrophysical Journal Supplement Series}, 250(2):27, sep
  2020.

\bibitem{Blasi_2012}
Pasquale Blasi and Elena Amato.
\newblock Diffusive propagation of cosmic rays from supernova remnants in the
  galaxy. i: spectrum and chemical composition.
\newblock {\em Journal of Cosmology and Astroparticle Physics},
  2012(01):010–010, January 2012.

\bibitem{Reichherzer_2022}
P.~{Reichherzer}, L.~{Merten}, J.~{D{\"o}rner}, J.~{Becker Tjus}, M.~J.
  {Pueschel}, and E.~G. {Zweibel}.
\newblock {Regimes of cosmic-ray diffusion in Galactic turbulence}.
\newblock {\em SN Applied Sciences}, 4:15, January 2022.

\bibitem{DiMauro_2023}
Mattia {Di Mauro}, Fiorenza {Donato}, Michael {Korsmeier}, Silvia {Manconi},
  and Luca {Orusa}.
\newblock {A novel prediction for secondary positrons and electrons in the
  Galaxy}.
\newblock {\em arXiv e-prints}, page arXiv:2304.01261, April 2023.

\bibitem{Xi_2019}
Shao-Qiang {Xi}, Ruo-Yu {Liu}, Zhi-Qiu {Huang}, Kun {Fang}, and Xiang-Yu
  {Wang}.
\newblock {GeV Observations of the Extended Pulsar Wind Nebulae Constrain the
  Pulsar Interpretations of the Cosmic-Ray Positron Excess}.
\newblock {\em \apj}, 878(2):104, June 2019.

\bibitem{Petrovic_2015}
Jovana {Petrovi{\'c}}, Pasquale~D. {Serpico}, and Gabrijela {Zaharijas}.
\newblock {Millisecond pulsars and the Galactic Center gamma-ray excess: the
  importance of luminosity function and secondary emission}.
\newblock {\em \jcap}, 2015(2):023--023, February 2015.

\bibitem{Macias_2021}
Oscar {Macias}, Harm {van Leijen}, Deheng {Song}, Shin'ichiro {Ando}, Shunsaku
  {Horiuchi}, and Roland~M. {Crocker}.
\newblock {Cherenkov Telescope Array sensitivity to the putative millisecond
  pulsar population responsible for the Galactic Centre excess}.
\newblock {\em \mnras}, 506(2):1741--1760, September 2021.

\bibitem{Ackermann_2015}
M.~{Ackermann} et~al.
\newblock {The Spectrum of Isotropic Diffuse Gamma-Ray Emission between 100 MeV
  and 820 GeV}.
\newblock {\em \apj}, 799(1):86, January 2015.

\bibitem{IGRB}
Mattia {Fornasa} and Miguel~A. {S{\'a}nchez-Conde}.
\newblock {The nature of the Diffuse Gamma-Ray Background}.
\newblock {\em \physrep}, 598:1--58, October 2015.

\bibitem{Pohl_2008}
Martin {Pohl}, Peter {Englmaier}, and Nicolai {Bissantz}.
\newblock {Three-Dimensional Distribution of Molecular Gas in the Barred Milky
  Way}.
\newblock {\em \apj}, 677(1):283--291, April 2008.

\bibitem{Su_2010}
Meng {Su}, Tracy~R. {Slatyer}, and Douglas~P. {Finkbeiner}.
\newblock {Giant Gamma-ray Bubbles from Fermi-LAT: Active Galactic Nucleus
  Activity or Bipolar Galactic Wind?}
\newblock {\em \apj}, 724(2):1044--1082, December 2010.

\bibitem{Ackermann_2014}
M.~{Ackermann} et~al.
\newblock {The Spectrum and Morphology of the Fermi Bubbles}.
\newblock {\em \apj}, 793(1):64, September 2014.

\bibitem{Tavakoli_2014}
Maryam Tavakoli, Ilias Cholis, Carmelo Evoli, and Piero Ullio.
\newblock Constraints on dark matter annihilations from diffuse gamma-ray
  emission in the galaxy.
\newblock {\em Journal of Cosmology and Astroparticle Physics}, 2014(01):017,
  jan 2014.

\bibitem{cirelli2015status}
Marco Cirelli.
\newblock Status of indirect (and direct) dark matter searches, 2015.

\bibitem{Martin_22}
Pierrick {Martin}, Luigi {Tibaldo}, Alexandre {Marcowith}, and Soheila
  {Abdollahi}.
\newblock {Population synthesis of pulsar wind nebulae and pulsar halos in the
  Milky Way. Predicted contributions to the very-high-energy sky}.
\newblock {\em \aap}, 666:A7, October 2022.

\bibitem{Catena_2010}
Riccardo {Catena} and Piero {Ullio}.
\newblock {A novel determination of the local dark matter density}.
\newblock {\em \jcap}, 2010(8):004, August 2010.

\bibitem{WechakamaAscasibar11}
M.~{Wechakama} and Y.~{Ascasibar}.
\newblock {Pressure from dark matter annihilation and the rotation curve of
  spiral galaxies}.
\newblock {\em \mnras}, 413(3):1991--2003, May 2011.

\bibitem{WechakamaAscasibar14}
M.~{Wechakama} and Y.~{Ascasibar}.
\newblock {Multimessenger constraints on dark matter annihilation into
  electron-positron pairs}.
\newblock {\em \mnras}, 439(1):566--587, March 2014.

\bibitem{constraints_Hess}
H.~{Abdalla} et~al.
\newblock {Search for Dark Matter Annihilation Signals in the H.E.S.S. Inner
  Galaxy Survey}.
\newblock {\em \prl}, 129(11):111101, September 2022.

\bibitem{Krommydas_2023}
Iason {Krommydas} and Ilias {Cholis}.
\newblock {Revisiting GeV-scale annihilating dark matter with the AMS-02
  positron fraction}.
\newblock {\em \prd}, 107(2):023003, January 2023.

\bibitem{Barak_2023}
Ramin {Barak}, Konstantin {Belotsky}, and Ekaterina {Shlepkina}.
\newblock {Proposition of FSR Photon Suppression Employing a Two-Positron Decay
  Dark Matter Model to Explain Positron Anomaly in Cosmic Rays}.
\newblock {\em Universe}, 9(8):370, August 2023.

\end{thebibliography}

\appendix
\section{SNR contribution to the leptonic gamma-ray emission}
\label{app:snr_gamma}

\begin{figure}
    \centering
    \includegraphics[width = 0.7\textwidth]{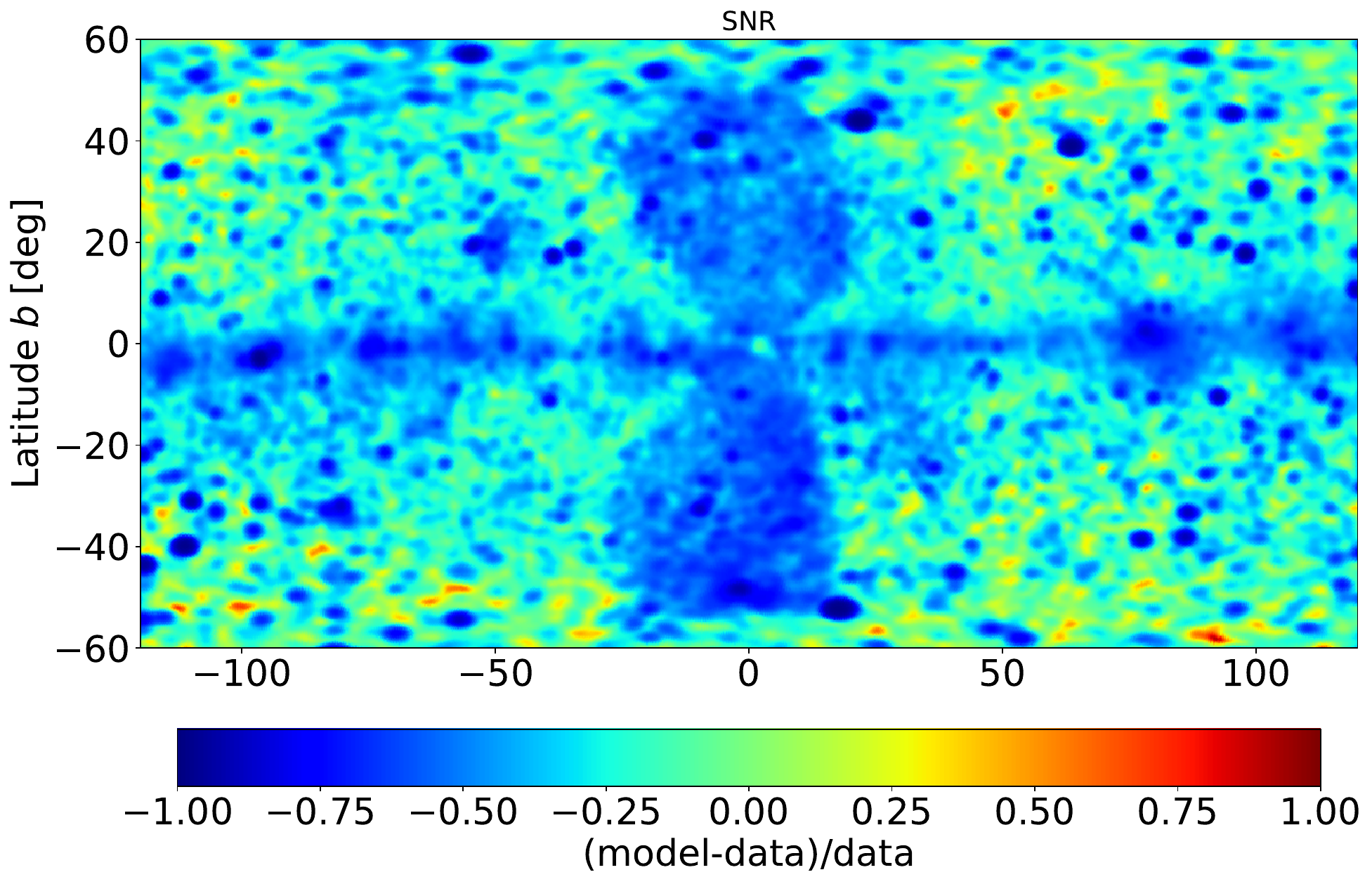}
    \caption{Residuals of the baseline SNR model.}
    \label{fig:snr_res}
\end{figure}

\begin{figure}
    \centering
    \includegraphics[width = \textwidth]{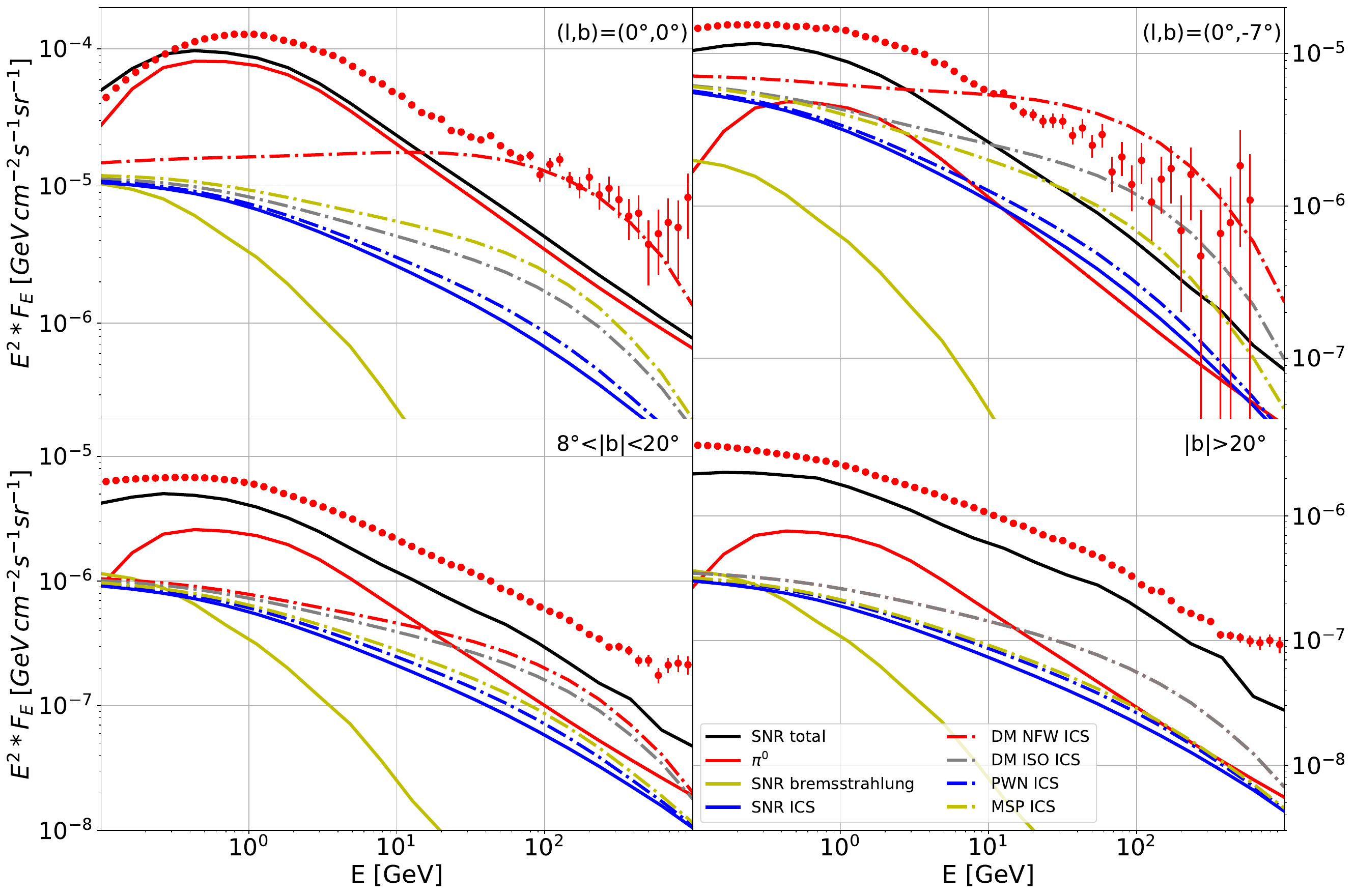}
    \caption{Gamma-ray spectra of the baseline SNR model, for different Galactic regions (specified on the top right of each panel). Its contribution from secondary electrons and positrons is depicted as solid blue (for ICS) and green (bremsstrahlung). The dash-dotted lines represent the ICS component of the rest of scenarios studied in this paper.}
    \label{fig:snr_gammaspec}
\end{figure}

As stated in Section~\ref{sec:gamma}, the $\pi^0$ component of the gamma-ray sky comes from the baseline SNR injection. However, this baseline model also contributes to the leptonic gamma-ray emission through its secondary electrons and positrons. The residuals of this baseline model and their contribution to the total gamma-ray emission can be found in Figure~\ref{fig:snr_res} and Figure~\ref{fig:snr_gammaspec}, respectively. The results are similar to those of the PWN scenario, since this profile does not inject many extra $e^-e^+$ pairs far from the Galactic plane.

\end{document}